\documentclass[nofootinbib, reprint, floatfix]{revtex4-1}

\usepackage{units}
\usepackage{graphicx} %Include figure files
\usepackage{upgreek}
\usepackage{mathtools}
\usepackage{dcolumn}% Align table columns on decimal point
\usepackage{bm}% bold math
\usepackage[caption=false]{subfig}
\usepackage{color,soul}
\usepackage{tabularx}
\usepackage{xcolor}
\usepackage{framed}

\begin{document}

%\preprint{APS/123-QED}

\title{Stimulated decay and formation of antihydrogen atoms}

\author{T. Wolz and C. Malbrunot}
\address{Physics Department, CERN, Gen\`eve 23, 1211, Switzerland}

\author{M. Vieille-Grosjean and D. Comparat}
\address{Laboratoire Aim\'{e} Cotton, CNRS, Universit\'e Paris-Sud, ENS Paris Saclay, Universit\'e Paris-Saclay, B\^{a}t. 505, 91405 Orsay, France}

\date{\today}

\begin{abstract}
Antihydrogen atoms are routinely formed at the Antiproton Decelerator at CERN in a wide range of Rydberg states. To perform precision measurements, experiments rely on ground state antimatter atoms which are currently obtained only after spontaneous decay. In order to enhance the number of atoms in ground state, we propose and assess the efficiency of different methods to stimulate their decay. First, we investigate the use of THz radiation to simultaneously couple all $n$-manifolds down to a low lying one with sufficiently fast spontaneous emission toward ground state. We further study a deexcitation scheme relying on state-mixing via microwave and/or THz light and a coupled (visible) deexcitation laser. We obtain close to unity ground state fractions within a few tens of \unit[]{$\mu$s} for a population initiated in the $n=30$ manifold. Finally, we study how the production of antihydrogen atoms via stimulated radiative recombination can favorably change the initial distribution of states and improve the overall number of ground-state atoms when combined with stimulated deexcitation.
\end{abstract}

%\pacs{Valid PACS appear here}

\maketitle

\section{\label{s:intro}Introduction}
After decades of technical developments, antihydrogen atoms are now routinely formed at CERN's Antiproton Decelerator (AD) complex \cite{bertsche2015physic}. The AD currently hosts five antihydrogen experiments aiming at precisely measuring physical properties of this anti-atom for stringent tests of the combined Charge-Parity-Time (CPT) symmetry and a first direct measurement of the effect of the gravitational force on antimatter.
In this quest, a plurality of experimental approaches has emerged. Anti-atoms are either trapped in magnetic fields for in-situ measurements \cite{AlP_Hbar_Accumulation,PhysRevLett.108.113002} or form a beam which is extracted away from the formation region into a quasi field-free environment \cite{kuroda2014source,KELLERBAUER2008351,Perez_2012}. In both cases, antihydrogen atoms in ground-state are needed to perform the intended measurements. They are however formed, in the vast majority of cases\footnote{The GBAR experiment, a new experiment at the AD, is relying on a process that should form quasi ground state atoms}, in highly excited states. 
Indeed, the main formation mechanisms are the so-called three-body-recombination (3BR) in which two positrons and an antiproton take part in the formation process (the additional positron carrying away the excess energy) and the so-called charge-exchange (CE) mechanism where a positronium atom (Ps:~a bound-state formed by an electron and a positron) in an excited state releases its positive charge to the antiproton and the remaining electron carries away the energy excess. The first mechanism is a ``quasi-continuous'' process which takes place as long as the positron and antiproton plasmas can be maintained in interaction (typically several hundreds of milliseconds \cite{kuroda2014source}) and produces a wide distribution of highly excited Rydberg atoms in all sub-states \cite{Gabrielse2002, rsa_Mal_18, ROB08,radics2014scaling, Jonsell_2019}. 
The CE mechanism can lead to a pulsed formation 
controlled by the laser-excitation time of the Ps atoms. In the CE mechanism, the distribution of the principal quantum number $n$ of the formed antihydrogen atoms is in part determined by the one of the Ps. Typically, experimental values around $n\sim30$ are targeted \cite{Krasnicky_2019,2016PhRvA..94b2714K,2012CQGra..29r4009D}, but with a wide distribution of substates. In summary, both formation processes form highly excited anti-atoms with a broad distribution of all ($l,m$) angular momenta. If we assume a statistical distribution of states, the high angular momentum states, which are the most populated levels, have radiative lifetimes toward ground-state of several tens of milliseconds. Experiments trapping antihydrogen atoms in magnetic traps can hold onto the atoms for much longer times and are thus able to gather ground-state atoms for measurements by spontaneous radiative decay \cite{ALP182,ALP181, ALP172}. Those measurements are however limited by the number of antihydrogen atoms which can be trapped owing to the large difference between their formation and trappable temperatures.  In contrast to trap experiments, those relying on a beam of antihydrogen atoms cannot afford to wait for spontaneous deexcitation of the formed antihydrogen atoms to perform the measurements. Even at state-of-the-art formation temperatures of $\sim$\unit[50]{K} \cite{AlP_Hbar_Accumulation}, yet to be demonstrated in a beam, typical velocities are of the order of $\unit[1000]{ms^{-1}}$ implying that an antihydrogen atom will travel several meters before reaching ground-state which leads to high losses via annihilations on the walls of the formation apparatus. Therefore, it is highly necessary that a stimulated deexcitation takes place at the moment of formation to quickly populate (in typically less than a few tens of $\unit[]{\mu s})$ the ground-state level. In a previous publication \cite{COM181} we have dealt with the case of pulsed deexcitation which can only be applied to pulsed CE formation. The proposed mechanism could achieve deexcitation to ground state in a sub-$\mu$s timescale, but suffered from a caveat that was overseen and later corrected \cite{PhysRevA.101.019904}. Here, we lay down the considerations which led to this correction and develop the case of a continuous deexcitation applicable also to the 3BR case.

As studied in Ref.~\cite{wetzels2006far}, the mere use of a laser to drive the deexcitation of antihydrogen atoms in a pure magnetic field is not efficient because it does not address the most-populated high angular momentum states. We thus propose in this work to first mix the states via THz and/or microwave radiation which leads to a reduced lifetime of the Rydberg atoms and also allows for dedicated stimulated deexcitation mechanisms to be implemented. A state-independent population transfer in Rydberg $^{85}$Rb atoms was demonstrated using half-cycle pulses \cite{mandal2010half} from $n=50$ to $n<40$ in view of application to antihydrogen. However, due to lack of powers and the fact  that the spectrum could not be shaped, the transfer was slow ($\unit[\sim 150]{\mu s}$) and not very efficient (only 10\% of the population deexcited).  

\begin{figure}
\centering
\includegraphics[width=1\linewidth]{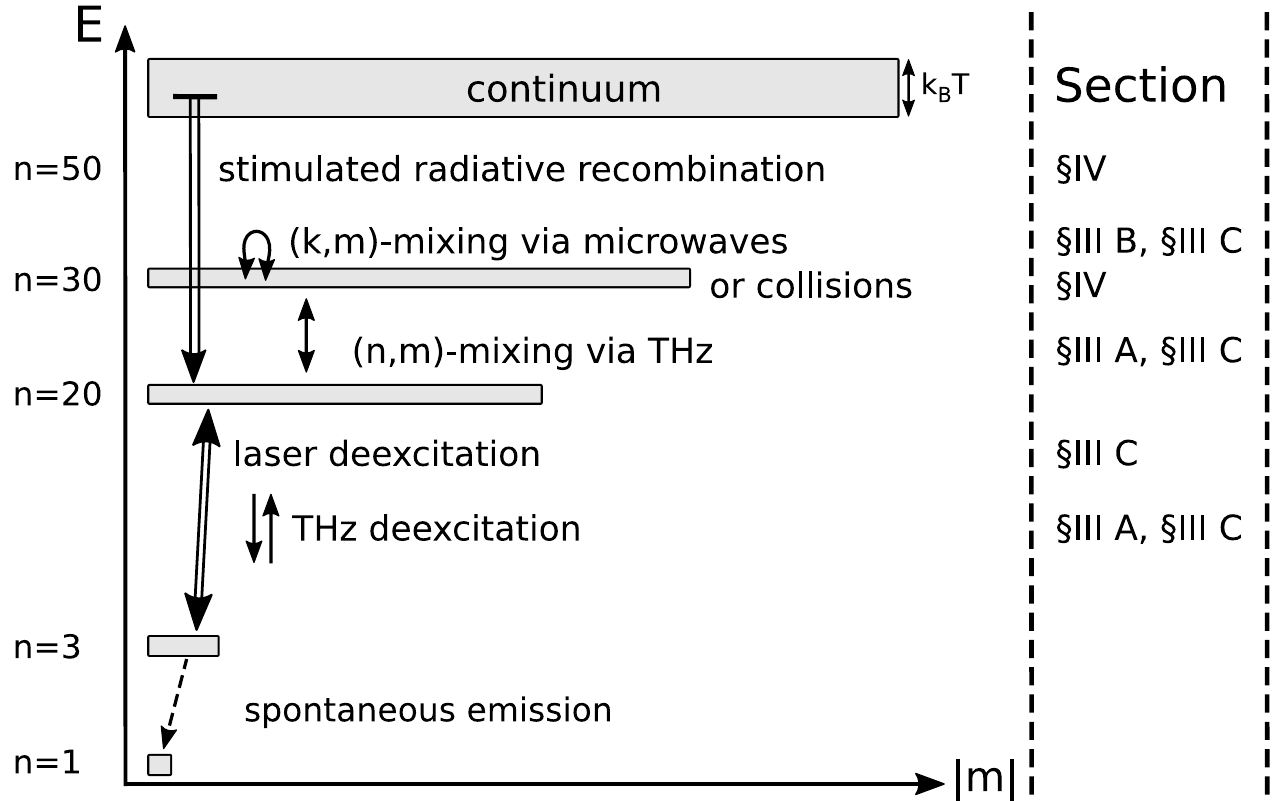}
\caption{Illustrative summary of the processes investigated. The dedicated sections are indicated on the right.}
\label{fig:IntroScheme}
\end{figure}

We first lay down in section \ref{s:physics} the important background considerations and assumptions made for the simulations of the atomic processes. We then investigate how to efficiently and rapidly bring atoms that have escaped the formation plasma to ground-state by stimulating inter-$n$-manifold atomic transitions toward low lying $n$ levels and intra-$n$-manifold transitions in the microwave frequency range. In both cases, the idea is to couple the distribution of states present after the antihydrogen formation to either low magnetic quantum number $m$ states that have short lifetimes and/or low $n$ manifolds exhibiting high spontaneous rates as well. The main principles are discussed using a generic model. The results of a full simulation are provided in section \ref{s: microwave, THz and laser}.
In section \ref{s:radrecomb} we discuss the benefit of combining a deexcitation scheme to an antihydrogen production via stimulated radiative recombination within the positron-antiproton plasma. Here, the coupling of states is achieved via collisions within the plasma.
We provide detailed studies of such process in a magnetic field. It is noteworthy to mention that section \ref{s: microwave, THz and laser} and \ref{s:radrecomb} stand on their own and can in principle be read independently.
A summary of all processes investigated and the associated sections in which they are discussed is given in Fig.~\ref{fig:IntroScheme}.

\section{\label{s:physics}General principles}

\subsection{General assumptions}\label{s:generalassum}

In the following, we will assume the antihydrogen atoms to be formed in a statistical distribution of states around the principal quantum number $n=30$ in a $\sim \unit[1]{T}$ magnetic field.
Typically, experiments relying on CE reaction will target such $n$ and magnetic field values, but the 3BR formation mechanism forms higher states in large quantities. However, in the standard experimental configurations, fields in \unit[]{few~kV/m} range at the edge of the antiproton plasma would ionize the antihydrogen atoms with roughly $n>50$  \cite{kuroda2014source,Gabrielse2002a,enomoto2010synthesis} so that outside of the few $\unit[]{mm}$ formation region, much higher Rydberg states are not present. 
The results presented in this manuscript are therefore valid for the proportion of the atoms formed around $n=30$ and lower, but the general considerations can be applied to higher states as an approximation as long as they fulfill the criterion $\left(\frac{n}{40}\right)^7 < \left(\frac{\bm B}{\unit[1]{T}}\right)^{-2}$, that is 
when the diamagnetic energy is small compared to the energy spacing between consecutive $n$-manifolds (i.~e.~below the $n$-mixing regime). In this inter-$l$-mixing regime other approximate quantum numbers (related to the Runge-Lenz vector) can be defined and $n$ can still be considered to be an approximately good quantum number \cite{friedrich1989hydrogen,cacciani1989anticrossing,pinard1990atoms}.
However, for higher $n$ states, excitation to higher manifolds and eventually ionization will play a larger role reducing the efficiency of the method proposed. We discuss in section \ref{s: Ionization_Excitation} the mechanisms at play.

In this manuscript, we treat the case of atoms in the presence of a pure magnetic field. In most experimental conditions however an additional electric field is present to hold the charged particles. This small additional field  (typically $\sim\unit[10]{V/cm}$) can lead to a perturbation of the states \cite{Cacciani1988, Braunand1984,Joerg1997}. A complete study of the combined magnetic and electric field effects is outside the scope of this paper, but in general it will create an additional mixing \cite{COM181} which is beneficial to the deexcitation goal, but will also induce potential losses through new excitation channels.

\begin{figure}
\centering
\includegraphics[width=1\linewidth]{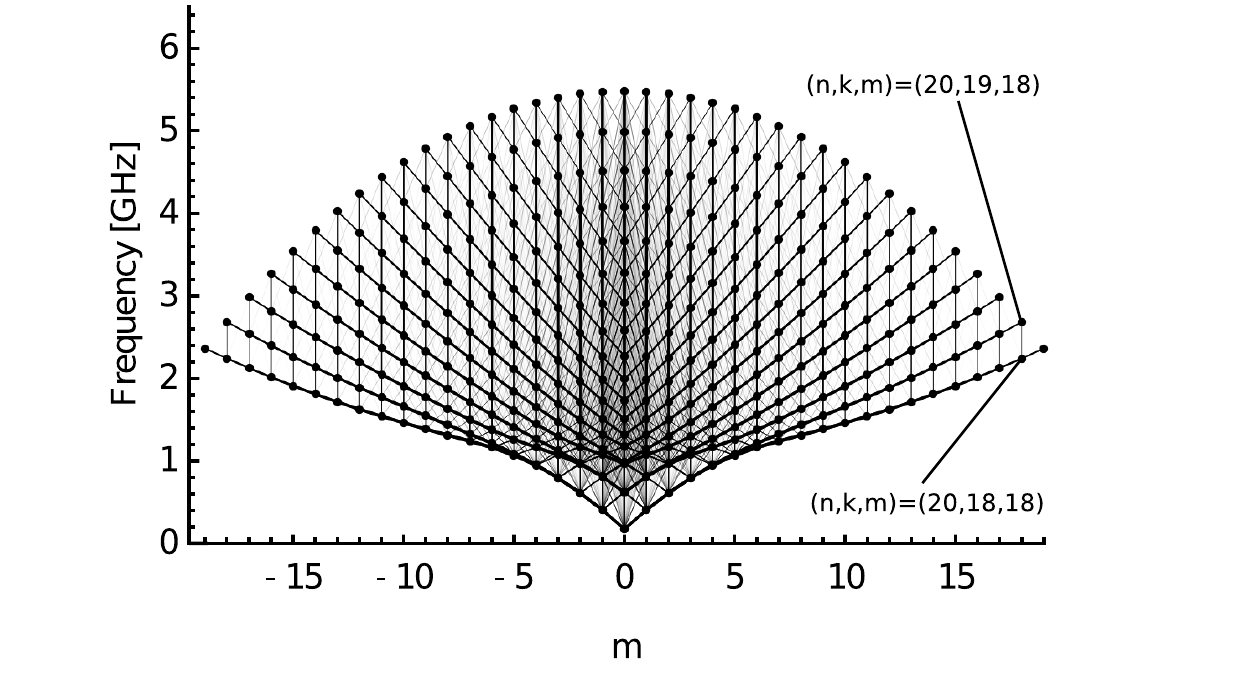}
\caption{Shift of energy levels  of all ($k,m$) sublevels (with respect to the $(k,m)=(0,0)$ state) of the $n=20$ manifold due to the sole diamagnetic term in a $\unit[1]{T}$ magnetic field. The thickness of the lines between states scales with the strength of the corresponding transition dipoles squared.}
\label{fig:cusp}
\end{figure}

In the presence of a pure magnetic field the magnetic quantum number $m$ (and parity) remains exactly defined since we neglect spin-orbit effects that are considered to be negligible for fields $\frac{B}{\unit[1]{T}}n^3 > 24$ \cite{friedrich1989hydrogen}. Thus, to obtain energy levels and transition dipoles from a state i to j as well as the required bandwidths to drive such transitions, we diagonalize the full Hamiltonian matrix 
\begin{equation}
	H_{\rm{ij}} = \left( E_{\rm{i}} + \frac{eB}{2 m_e c} m \right) \delta_{\rm{ij}} + \frac{e^2 B^2}{8 m_e c^2} H^{\rm{Q}}_{\rm{ij}}
\label{eq:Hamiltonian}
\end{equation}
for each set of ($n,m$) states. $E_{\rm{i}}$ is the zero-field energy and the matrix elements $H^{\rm{Q}}_{\rm{ij}}=\langle {\rm{i}} | r^2 {\rm{sin}}^2 \theta | {\rm{j}} \rangle$ are given in usual polar coordinates $r$ and $\theta$ by \cite{gallagher1994,Garstang1974}.
The field-free $l$ states are now labeled by an index $k$ (with $|m| \leq k < n$) according to the magnitude of diamagnetic interaction.
The fact that $m$ is a good quantum number is important because we can investigate $m$-manifolds separately. For instance, we can separate the linear Zeeman effect from the diamagnetic term and thus illustrate the energy spread of $(k,m)$ sublevels due to the sole diamagnetic term. Fig.~\ref{fig:cusp} indicates, for example, the diamagnetic spread for the $n=20$ manifold where transitions between the states are illustrated as lines between the levels with a thickness that scales with the strength of the dipole squared. Since in a magnetic field $l$ is not a good quantum number anymore, the only selection rule is $\Delta m = 0,\pm1$ which results in a mixing of angular momenta visible primarily at low $m$.

Given the broadband and thus incoherent nature of the light used for stimulating the decay, we use rate equations to simulate the atomic processes. Given the short timescales involved in the stimulated deexcitation, spontaneous emission from the long lived Rydberg levels coupled to each other via THz and/or microwave is very small. This prevents the formation of dark-states via spontaneous emission \cite{auzinsh2008, lindvall2013}. Further, we argue that dark state formation due to locally linear or circular polarized radiation \cite{PhysRevA.65.033413} would be limited, from an experimental point of view, by the light reflections on the metallic surfaces inside the experiment as well as by the use of  multiple light sources.

As motivated in section \ref{s: Ionization_Excitation}, ionization through off-resonance transitions generated by the THz, microwave or laser light used for deexcitation is small in the range of powers and frequencies used, but excitation to $n>30$ states has a significant impact on the results.
From section \ref{s:physics} to \ref{s: microwave, THz and laser} we consider antihydrogen atoms which have left the formation region and are therefore not anymore subject to collisions with the dense plasma of antiprotons and positrons. In section \ref{s:radrecomb} we will treat the case of formation and deexcitation within the plasma.

\subsection{Spontaneous decay}
In beam experiments, it is reasonable to target a deexcitation to ground-state in a few tens of $\unit[]{\mu s}$ so that the atoms formed in a typical cloud size of \unit[1]{mm} and average velocities of $\sim\unit[1000]{ms^{-1}}$ do not expand more than the original charged particle trap's size (typically of a few \unit[]{cm} radius) before they reach ground-state.

In a field-free region the spontaneous lifetime of a ($n,l,m$) state (with $|m| \leq l < n$) can be approximated by \cite{PhysRevA.31.495} 
\begin{equation}
		\tau_{\rm{n,l}} \approx \left( \frac{n}{30} \right)^3 \left( \frac{l+1/2}{30} \right)^2 \times \unit[2.4]{ms}.
\label{eq: RydbergLifetime}
\end{equation}
The result is also a good approximation for the presently treated case with $n \sim 30$ and in a \unit[1]{T} magnetic field environment \cite{Topccu2006} showing that for  high ($n,m$) states, the spontaneous lifetime is several orders of magnitude too high to allow for a rapid enough population of the ground-state following the atoms' formation.

\subsection{Energy levels and transitions in Rydberg antihydrogen}
\label{s: energy levels and transition freq}
\begin{figure}
\centering
\includegraphics[width=1\linewidth]{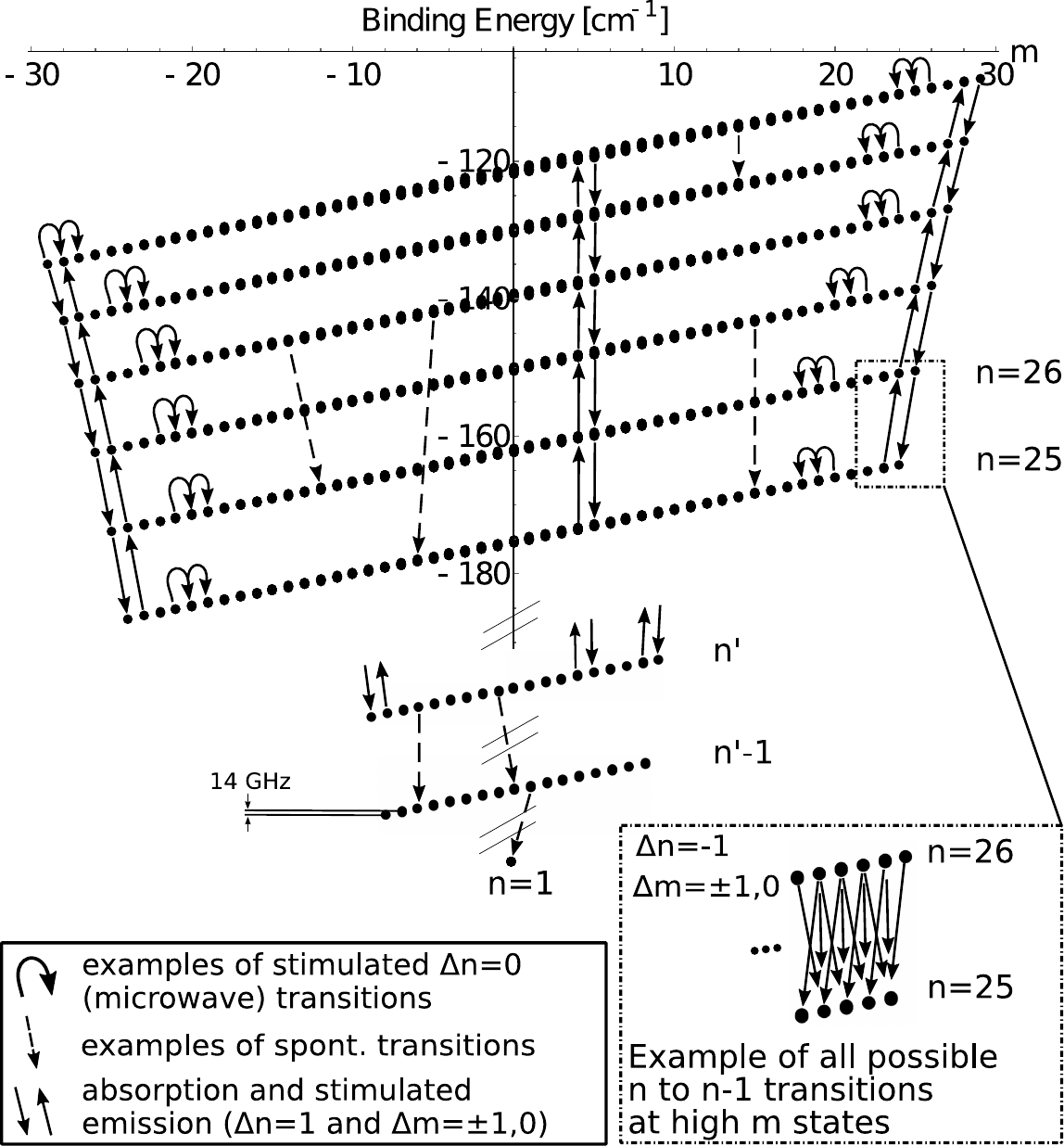}
\caption{Transitions and binding energy diagram of hydrogen levels as a function of the magnetic quantum number $m$ in a \unit[1]{T} magnetic field. The stimulated $\Delta m =0,\pm 1$ transitions  are represented by continuous arrows. Microwave transitions with $\Delta n =0$ are indicated by curly arrows. Some examples of spontaneous decays are indicated by dashed arrows.}
\label{fig:THz_deexcitation} 
\end{figure}

\begin{figure}
\centering
\subfloat[Inter-$n$-manifold transition frequencies and bandwidths\label{fig:2a}]{\includegraphics[width=1\linewidth]{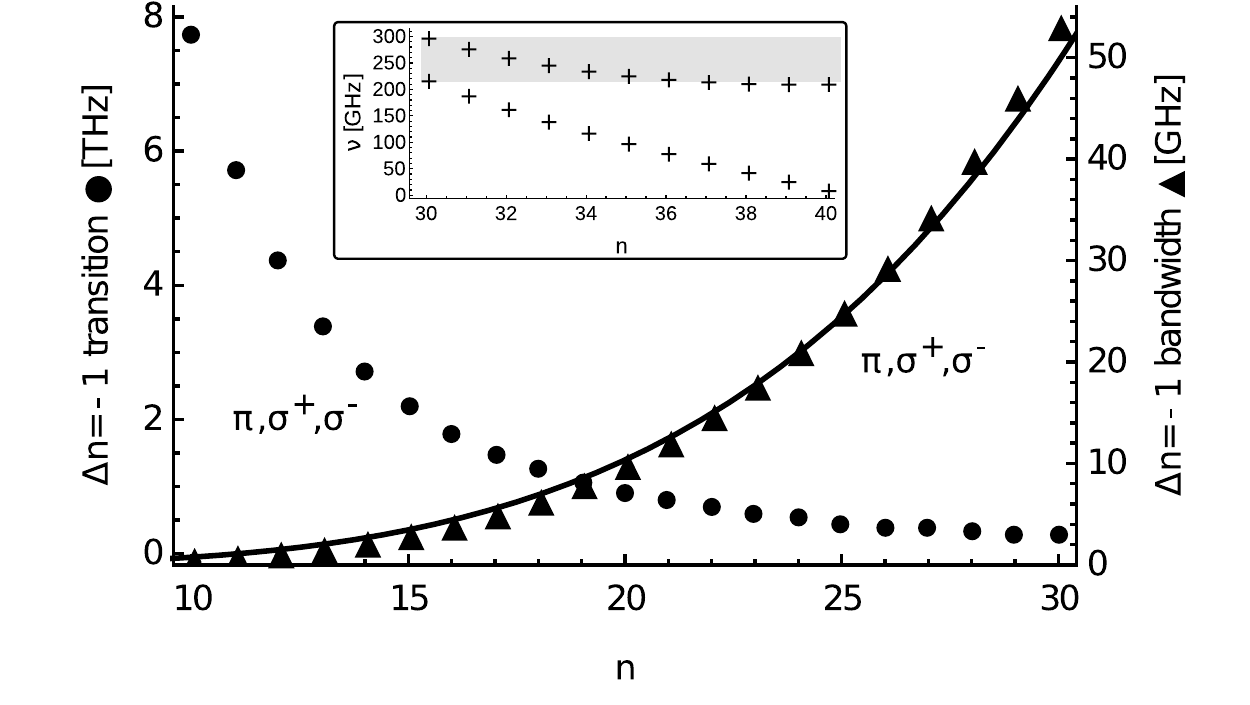}}\newline
\subfloat[Intra-$n$-manifold transition bandwidths\label{fig:2b}]{\includegraphics[width=1\linewidth]{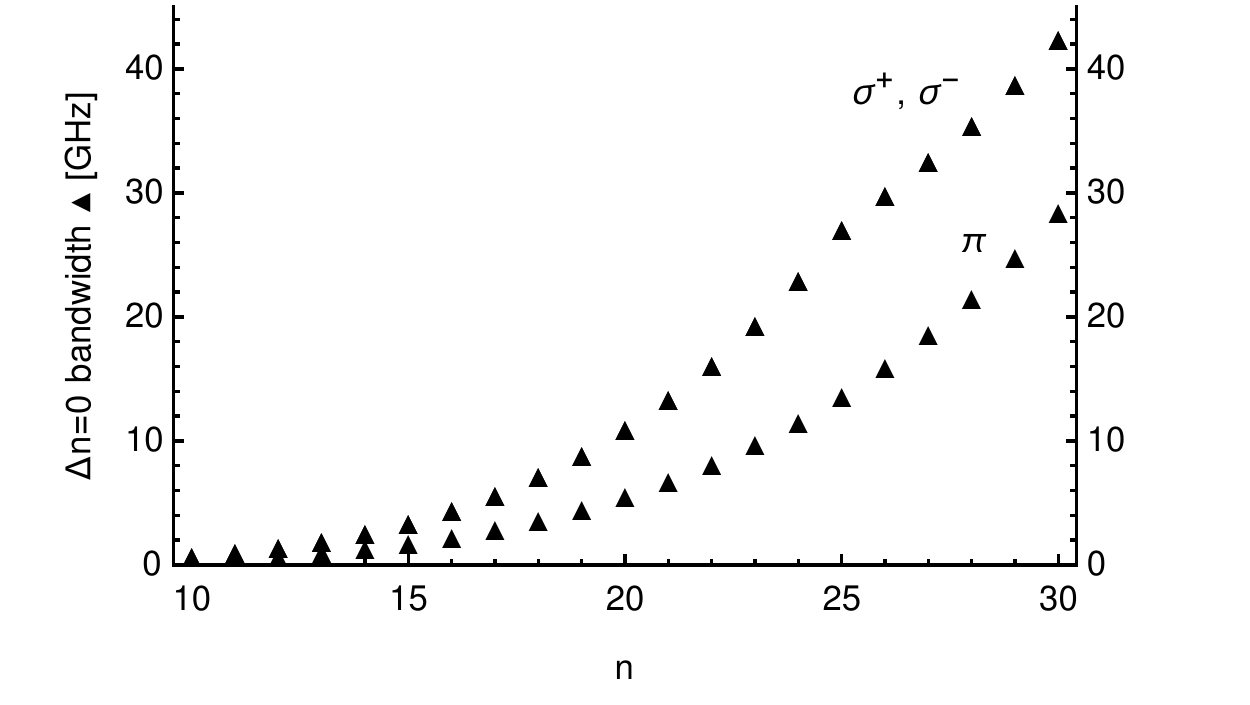}}
\caption{ (a) Hydrogen $\Delta n =-1$  transition frequencies (left axis) as well as the $\pi$, $\sigma^+$ and $\sigma^-$ transition bandwidths (right axis) as a function of $n$ in a \unit[1]{T} magnetic field. The fit to the bandwidth dependence on $n$ gives $ \unit[6.5\times10^{-5}]{GHz} \times n^4$. The transition frequencies are indicated as being the central frequency of all present transitions for a given polarization (given the scale of the axis, the $\pi$ and $\sigma^{\pm}$ curves cannot be distinguished from each other). The inset shows the minimum and maximum transition frequencies for $n$ to $n-1$ transitions from $n=30$ to $n=40$, taking into account all polarizations, in a $\unit[1]{T}$ magnetic field, see section \ref{s: Ionization_Excitation}. The shaded area illustrates the bandwidth of the $n=30$ to $n=29$ transitions and its overlap with the bandwidth of transitions from $n>30$. (b) Hydrogen $\Delta n =0$  transition bandwidths for all three polarizations.}
\label{fig:FreqBandwidth}
\end{figure}

Fig.~\ref{fig:THz_deexcitation} shows the binding energy of antihydrogen Rydberg levels as a function of $m$ in a magnetic field of \unit[1]{T}. Given the electric dipole transition selection rules, only transitions with $\Delta m = 0,\pm1$ are allowed. That implies that for states with maximal magnetic moments, the only possible inter-manifold transitions toward ground-state are those with $\Delta n =-1$ meaning that all such transitions are necessarily involved when decaying to ground state (cf.~ section \ref{s:THz}). The frequencies of these transitions (cf.~Fig.~\ref{fig:2a}) range for linearly polarized light from over \unit[7.5]{THz} for $n=10\rightarrow 9$ to \unit[0.26]{THz} for $n=30 \rightarrow 29$. The $\sigma^\pm$-transition frequencies are detuned from these values by approximately $\unit[\pm 14]{GHz/T}$, which is the linear Zeeman shift $\mu_B B$ for $\Delta m=\pm 1$, where $\mu_B$ is the Bohr magneton. The triangle markers in Fig.~\ref{fig:2a} indicate the $n\rightarrow n-1$ transition bandwidths in order to ensure a coverage of the energy shift of all addressed $k$ states for a given light polarization. In a \unit[1]{T} magnetic field this leads to linewidths from $\unit[0.5]{GHz}$ for $n=10$ to $\unit[53.1]{GHz}$ for $n=30$.

Intra-manifold $\Delta n=0$ transitions are indicated in Fig.~\ref{fig:THz_deexcitation} by curly arrows. Among those, the $\Delta m=0$ transitions are purely diamagnetic thus the transition frequencies are small and the bandwidth of the light to address all $\Delta m=0$ transitions within a given $n$-manifold is given by the spread of the ($k,m$) sublevels corresponding for $n=20$, for example, to \unit[5.6]{GHz} (cf.~Fig.~\ref{fig:cusp}). For $\Delta m=\pm1$, the linear Zeeman term ($\sim \unit[14]{GHz}$) adds up to the transition frequency and the bandwidth equals roughly twice the $\Delta m=0$ bandwidth until $n\sim 25$ where the diamagnetic shift starts to exceed $\mu B$, resulting in a constant frequency shift of \unit[14]{GHz} between the $\pi$ and $\sigma^{\pm}$ polarizations (cf.~Fig.~\ref{fig:2b}).   
For $\pi$ ($\sigma^\pm$) polarized light, the bandwidths range from \unit[0.31]{GHz} (\unit[0.62]{GHz}) for $n=10$ to \unit[28.4]{GHz} (\unit[42.4]{GHz}) for $n=30$. 

\subsection{Excitation and ionization}
\label{s: Ionization_Excitation}
Excitation and ionization processes set a limit to the light frequencies and corresponding intensities which can be used to deexcite the cloud of antihydrogen atoms.

The light used to drive $\Delta n= -1$ transitions can lead to excitation to higher levels. The insert in Fig.~\ref{fig:2a} shows the coverage of the $30\rightarrow31$, $31\rightarrow32$ etc.~transitions by the bandwidth (illustrated by the gray area between the minimum and maximum frequencies required to drive the transitions between the $n=30$ and $n=29$ manifolds) of the light used to drive the $30\rightarrow 29$ transitions. It illustrates that light with $\sim\unit[50]{GHz}$ spectral linewidth at $\sim\unit[0.26]{THz}$ can potentially drive population up to $n=35$. With the same mechanism, contributions from the $29\rightarrow28$, $28\rightarrow27$ etc.~transitions will also add up, but given that the bandwidth gets smaller and the level spacing gets larger this effect becomes rapidly negligible so that it is sufficient to illustrate it with the $30\rightarrow29$ transitions. On the contrary, the effect becomes rapidly larger for higher $n$. This shows that $n\sim30$ is close to an optimum which maximizes the number of states addressed in the initial distribution of antihydrogen while keeping the losses via excitation at a reasonable level. In the simulations presented in section \ref{s: microwave, THz and laser}, we consider $n=30$ as the highest state targeted. The effect of excitation up to $n=35$ levels is always taken into account in the simulation and mainly leads to a longer deexcitation time compared to when excitation is neglected.

THz frequencies necessary to drive low $\Delta n = -1$ transitions can couple high $n$-manifolds to the continuum. The ionization thresholds for antihydrogen Rydberg atoms lie around $\unit[3.7]{THz}$ for $n=30$, $\unit[2.1]{THz}$ for $n=40$ and $\unit[1.3]{THz}$ for $n=50$ corresponding roughly to $\Delta n =-1 $ transition frequencies of $n=13$, $15$ and $20$. 
To evaluate the effect of photoionization we use, for a given $n$-manifold, the extra photon energy $E = \kappa^2 R_y = \hbar \omega - R_y/n^2$ above the ionization threshold where $R_y$ is the Rydberg energy and calculate the photoionization cross-sections.
We use the field-free wavefunctions for the continuum which can be justified by the fact that the thermal spread  $\sim k_B T$ is larger than the energy of the cyclotron frequency $\hbar e B/m$. Thus, this tends to smear out the Landau quantization of the cyclotron frequency in the continuum (the bottleneck arises at a temperature below \unit[1.3]{K} per Tesla) \cite{ROB08}. 
Furthermore, for this estimation of the photoionization effect we use an averaged cross-section, assuming unpolarized light, defined by $\sigma_{n,m}^{\kappa} = \sum_{l=|m|}^{n-1}\frac{\sigma_{n,l}^{\kappa,l+1}+\sigma_{n,l}^{\kappa,l-1}}{2l+1}$ from each ($n,m$) level toward the continuum. This is similar to 
the assumption of a full $k$ (or $l$) mixing (as done in the appendix section \ref{sec: Srr and photoionization rates}).
Formulae for $\sigma_{n,l}^{\kappa,l'}$ are given in \cite{COM181}. We implement ionization rates $\Gamma_{n,m}^{\kappa} = \frac{I}{\hbar \omega} \sigma_{n,m}^{\kappa}$ for the entire set of intensities $I$ and frequencies $\omega$ used (all the ones driving $\Delta n=-1$ transitions from $n=30$ down to low $n$ states).
We find that, in the treated case, the ionization does not play a significant role even when using a total light intensity of $\unit[500]{W/m^2}$ (corresponding to the highest total intensity used in the simulation) which is comparable to the intensity emitted by a blackbody source at \unit[300]{K}. Ref.~\cite{Seiler_2016} studied the effect of a \unit[300]{K} blackbody spectrum on hydrogen in $n=30$ and $0 \leq | m | \leq 15$ states and showed that $\sim\unit[15]{\%}$ of the atomic sample is ionized in $\unit[100]{\mu s}$, a result also confirmed in \cite{vieil2018}. We propose to use comparable intensities as present in the THz part of the \unit[300]{K} blackbody spectrum to drive low $n$ transitions. However, the timescales involved in the work presented here are significantly shorter (few tens of $\unit[]{\mu s}$ instead of $\unit[100]{\mu s}$). Additionally, during this time, the population is rapidly driven toward low $n$ states that couple less strongly to the continuum than the $n=30$ states. Finally, treating states with higher $m$ values, for which the ionization cross-section is small, contributes as well to minimizing the ionization rate.
In the following simulations we thus take into account the ionization process, but, as expected from the above considerations, it does not significantly impact the results presented.

Frequencies in the microwave region to stimulate $\Delta n=0$ transitions can also lead to photoionization. Microwave ionization can be a complex process with multiple (non-) adiabatic crossings, multi-photon processes, Anderson's localization scenario, etc.~\cite{koch1995importance,krug2005universal}. It is therefore out of the scope of this article to study this process in detail. However, given the ratio of the $n$ to $n+1$ level spacing and the microwave frequencies (and bandwidths) required to drive the intra-manifold transitions, several photons would be required to excite to higher manifolds which makes it a negligible process. Another coupling to the continuum  can be achieved by cascade if a strong electric field couples $n$ and $n+1$ levels by Stark mixing \cite{gallagher1994}. With such a field, all other higher levels will also be coupled culminating in ionization.
In a magnetic field-free environment, a field near the Inglis-Teller one $\unit[7]{kV/cm}(n/30)^{-5}$ would thus be sufficient to ionize the atoms. This would correspond for $n=30$ to a microwave power of $\sim \unit[10^5]{W/m^2}$  which is several orders of magnitude higher than the needed intensity discussed in section \ref{s:RF} and \ref{s : laser}. Even though the limit will be somewhat lower in a magnetic field because the $n$ and $n+1$ manifolds are closer in energy to each other, we neglect the effect of multi-photon processes and ionization by microwave radiation for the mixing mechanism studied here and additionally note that we will illustrate the mechanism using microwave sources covering the entire spread caused by the diamagnetic term (cf.~Fig.~\ref{fig:cusp}), but that several optimizations can be done based on light polarization or the selection of a small number of states.  

\subsection{Generic model}

We propose in this manuscript to couple a large number of Rydberg quantum states (typically a few thousand) to fast spontaneously decaying levels. The key points when driving several states to few short-lived final ones is that the population transfer between the states has to happen faster than the dissipative process (that is the spontaneous decay) and ideally at the same rate since the slowest rate will always constitute a bottleneck. 
It is thus desirable to establish equal stimulated rates in order to reach a steady state (i.e. equipopulate all involved levels) as fast as possible. Coupling an initial population of $N$ Rydberg states that have lifetimes of the order of $\tau_{N} \sim \unit[]{ms}$ to a number of target levels $N'$ with an average deexcitation time to ground state of $t^{\rm{GS}}_{N'} \ll \tau_{N}$ leads to a total deexcitation time $t_{\rm{deex}}$ of
\begin{equation}
\label{eq: DeexcitationScal}
    t_{\rm{deex}} = \frac{N}{N'} \times t^{\rm{GS}}_{N'}.
\end{equation}
It is implicitly assumed here that the cascade from $n$ to $n'$ is dominated by the stimulated channel with spontaneous rates neglected. Despite this approximation, Eq.~\ref{eq: DeexcitationScal} gives a very helpful first insight into the characteristic deexcitation time.
The estimation of the time to decay to ground state $t^{\rm{GS}}_{N'}$ is not obvious because of the many involved exponential decays which lead to different behaviors at short and long times. Considering low lying $n'$ manifolds, we can approximate this time by the lifetime of the given $n'$ manifold. Averaging the decay times found in Eq.~\ref{eq: RydbergLifetime} results in: $ t^{\rm{GS}}_{N'} = \frac{1}{n'^2} \sum_{l'=0}^{n'-1} \left( 2l'+1 \right) \tau_{\rm{n',l'}} \approx \unit[5]{\mu s} \times (n'/10)^5$ which is an over-estimation due to the dominance of long-lived circular states. If instead we assume the ($k,m$) sub-levels to be mixed and average on the rates we find a scaling $t^{\rm{GS}}_{N'} \approx \unit[2]{\mu s} \times (n'/10)^{4.5}$ \cite{COM181}. Both approaches yield similar results in particular for low $n'$ and approximate well, in the range of $n$ states considered, the analytical result found in \cite{PhysRevA.31.495}.   
Thus, if, for example, the $20 \leq n \leq 30 \sim 7000$ levels are coupled to the $n'=3$ manifold that decays to ground state in $\sim \unit[10]{ns}$, the characteristic deexcitation time will be of the order of $t_{\rm{deex}} \sim \unit[8]{\mu s}$. This has to be compared to the $\sim \unit[100]{ns}$ which were found in \cite{COM181} using a single broadband laser driving the $20 \leq n \leq 30$ population to $n'=3$. There, it was assumed that the initial distribution of states which was fully ($m,k$) mixed by an appropriate choice of electric and magnetic field values and relative orientation could be treated as a single steady state. The considerations above show that this is not a valid assumption.
In order to correct the results of Ref.~\cite{COM181}, we solved the set of rate equations for all mixed ($n,m_1,m_2$) levels with $20\leq n \leq 30$ under the presence of the optimum electric and magnetic field values found in \cite{COM181} and implement laser stimulated rates to the $n'=3$ manifold that decays spontaneously to ground state. We found that $\sim 60\%$ of the atoms with an initial statistical distribution in the $20\leq n \leq 30$ manifolds are brought to ground state within $\unit[10]{\mu s}$ which is in good agreement with the considerations and the generic model presented above, see also \cite{PhysRevA.101.019904}. 

\section{Stimulated deexcitation and mixing}	
\label{s: microwave, THz and laser}

\subsection{\label{s:THz} THz stimulated deexcitation and mixing}
The first investigated technique to accelerate the decay toward the ground state consists in using THz light to stimulate all $ \Delta n = \pm 1$ transitions between an initial $n$ manifold down to a manifold $n'$ from where the spontaneous emission is fast enough. As mentioned earlier, driving these transitions allows to address high $m$ states for which $\Delta n = - 1$ are the only possible transitions toward lower states. These high angular momentum states are particularly important in the context of a fast stimulated deexcitation since they are incidentally the states with the longest lifetimes and highest population probability.

Since the exact distribution of states is experimentally not known \cite{rsa_Mal_18,Gabrielse2002}, the choice of an initial distribution to present the simulation results can be somewhat arbitrary.  However, it is worth noting that, in the deexcitation process, all transitions $n\rightarrow n-1$ of the cascade have to be driven simultaneously (and not sequentially) in order to avoid a mere population exchange between the levels. Therefore, the fact that all such frequencies are present, allows us to estimate the efficiency of the processes by 
restricting the initial distribution to a single manifold (typically $n=30$) to better highlight the dynamics of the THz-induced deexcitation and mixing mechanisms. As mentioned previously, $n=30$ was found to be close to the highest state which can efficiently be targeted with this method. In the case of a broader initial population below $n=30$, the performance will, in general, improve because there is no need to deexcite the population anymore. In such a case the states need to be merely coupled to each other to retain equipopulation.

In order to extract the power needed for the deexcitation and the efficiency of the method, we simulate the atomic system under consideration by implementing all spontaneous and stimulated rates between all ($n,k,m$) levels into a complex set of rate equations. The results presented in this section are obtained by implementing, all types of radiation (if multiple are present, cf.~section \ref{s:RF} and \ref{s : laser}) simultaneously into the equation system. We always solve the rate equations for states up to the $n=35$ manifold (motivated in section \ref{s: Ionization_Excitation}) in order to take into account excitation processes. The simulation considers unpolarized light ($\frac{1}{3} \sigma^+$, $\frac{1}{3} \sigma^-$ and $\frac{1}{3} \pi$). The presence of different polarizations is key since, for example, under pure $\pi$ polarized light, no $m$ mixing would occur and thus states with high $m$ values would never be stimulated to decay.

The dynamic of the simulation is not only influenced by the THz power used but also by the $n$ dependence of the power required to drive the $n\rightarrow n-1$ transition, that is, how the total power is distributed among the transitions addressed. We simulated three different scalings: a \emph{flat} scaling corresponding to an equal distribution of the power among the different transitions, a \emph{linear increase} scaling where more power is distributed to low $n$ with a slope such that the final $n'+1\rightarrow n'$ transitions are driven with an intensity $100$ times stronger than the initial $n= 30\rightarrow29$ transitions, and finally a \emph{linear decrease} where higher power is distributed to high $n$ with the same slope as for the previous scaling.

\begin{figure}
\centering
\subfloat[\label{fig:THzPlot1}]{\includegraphics[width=0.5\linewidth]{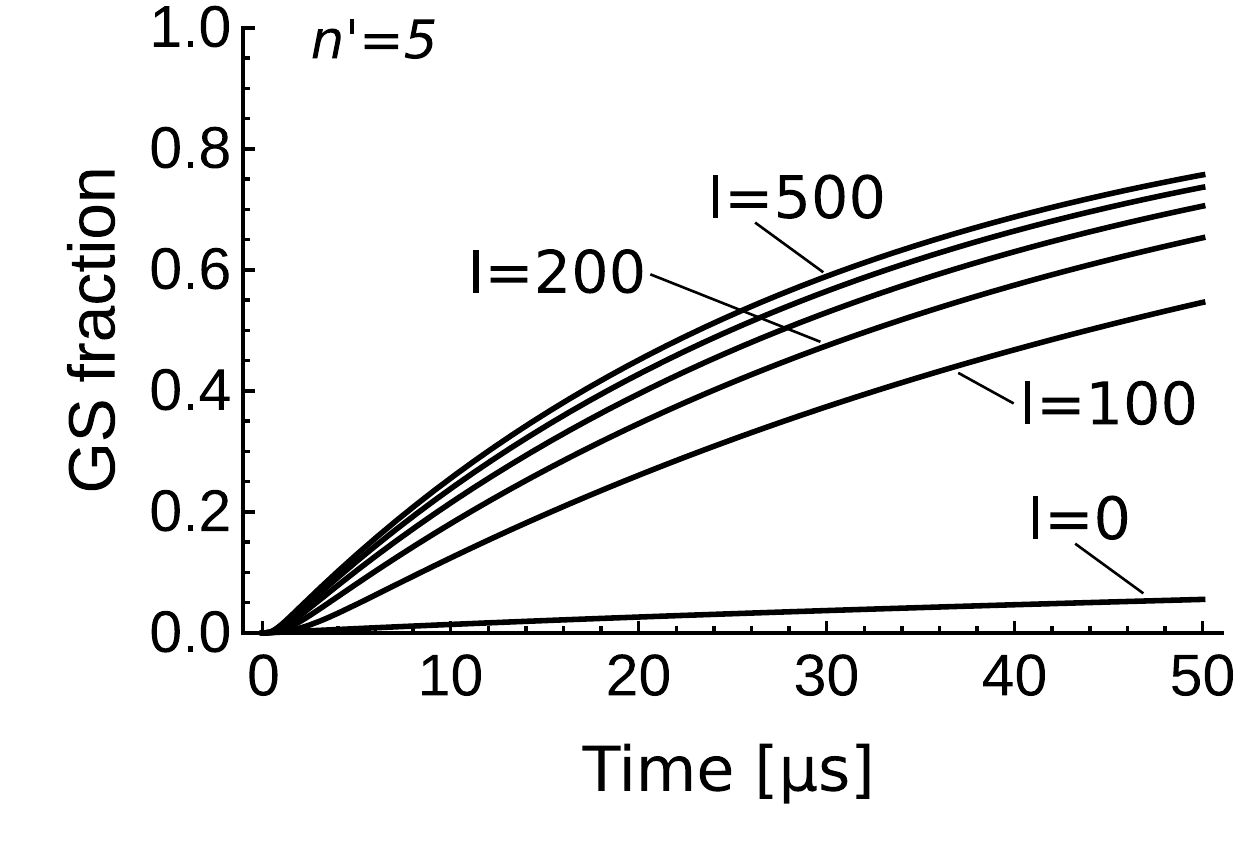}}
\subfloat[\label{fig:THzPlot2}]{\includegraphics[width=0.5\linewidth]{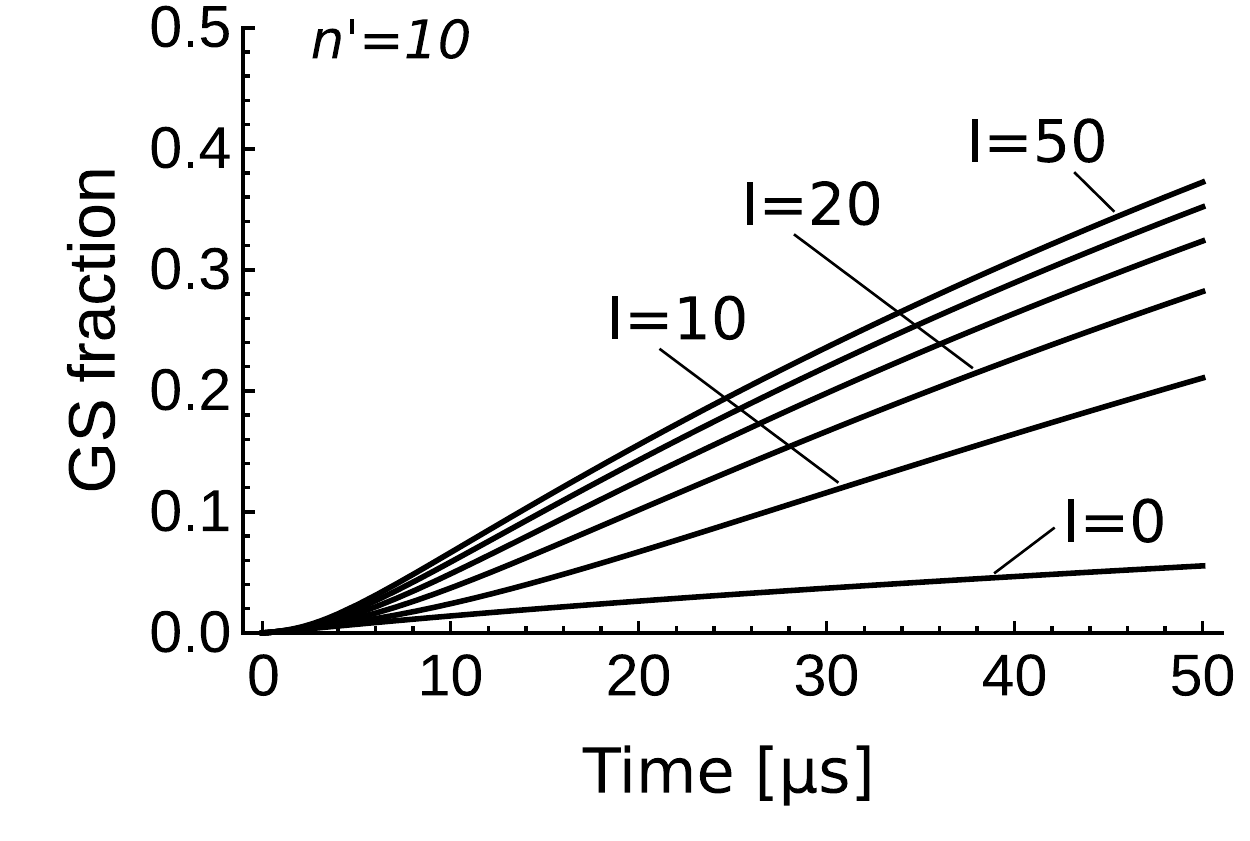}}

\subfloat[\label{fig:THzPlot3}]{\includegraphics[width=0.5\linewidth]{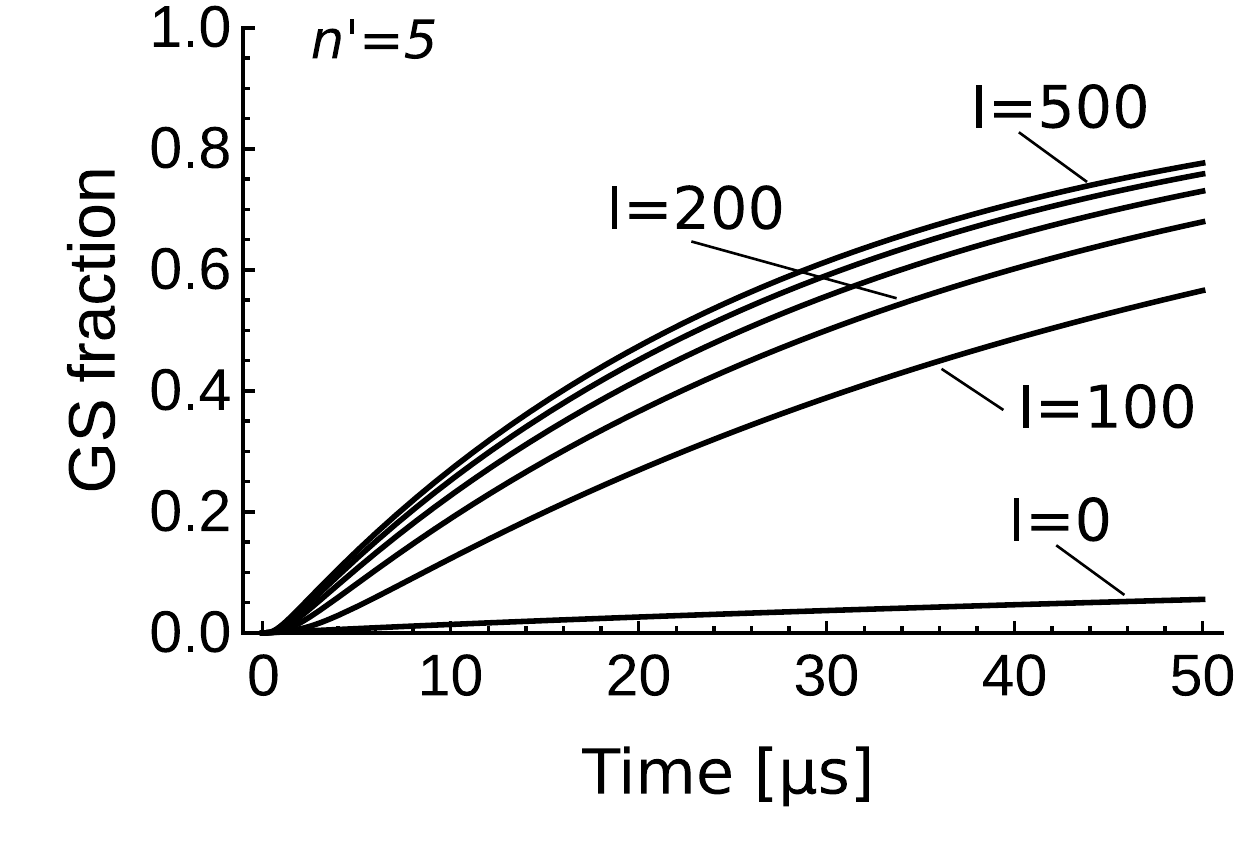}}
\subfloat[\label{fig:THzPlot4}]{\includegraphics[width=0.5\linewidth]{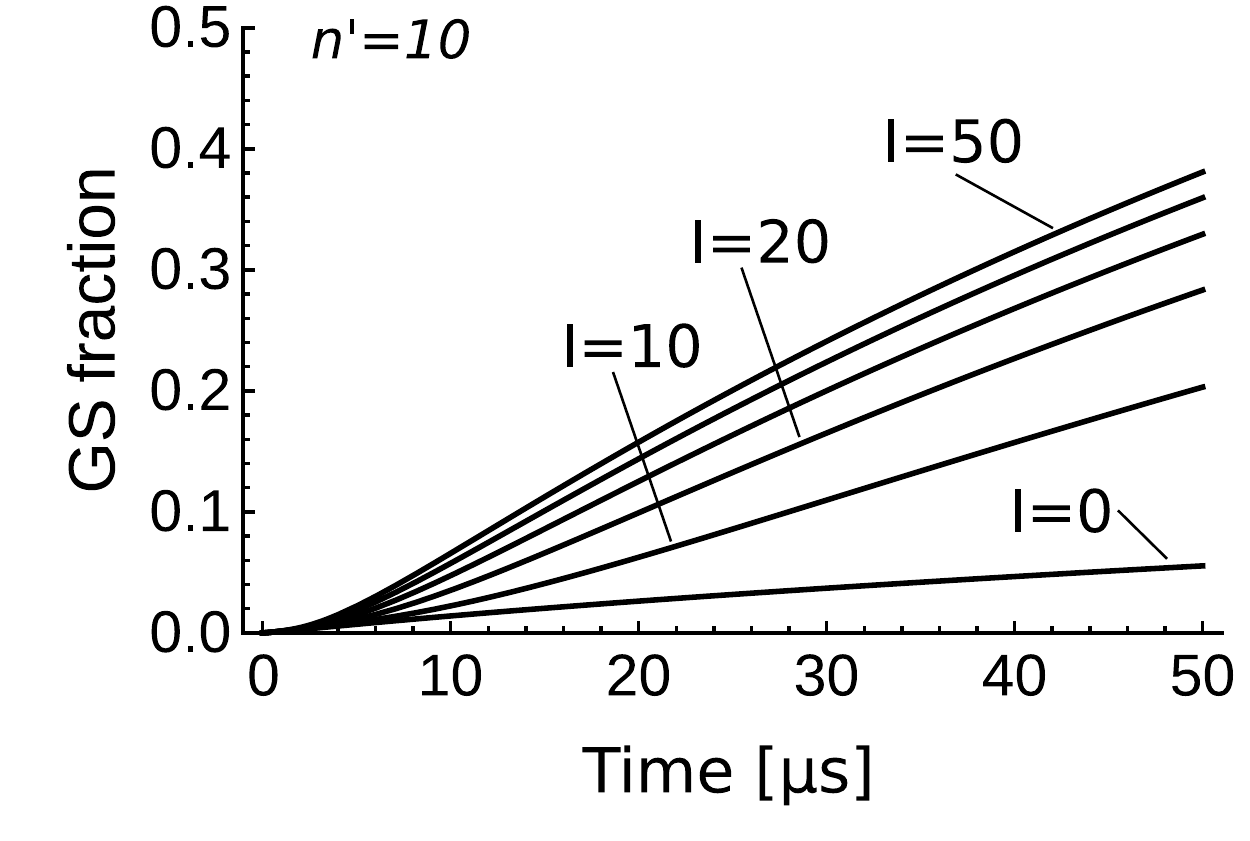}}

\subfloat[\label{fig:THzPlot5}]{\includegraphics[width=0.5\linewidth]{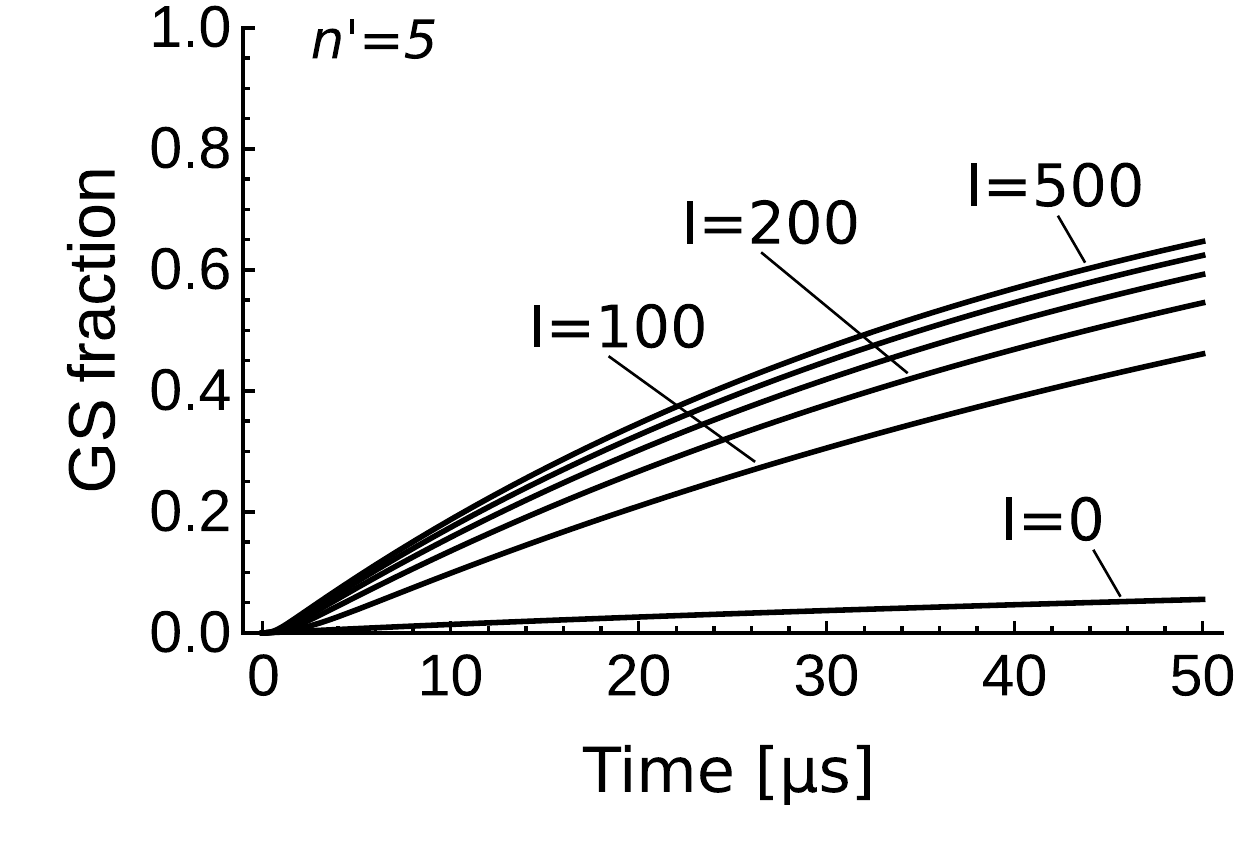}}
\subfloat[\label{fig:THzPlot6}]{\includegraphics[width=0.5\linewidth]{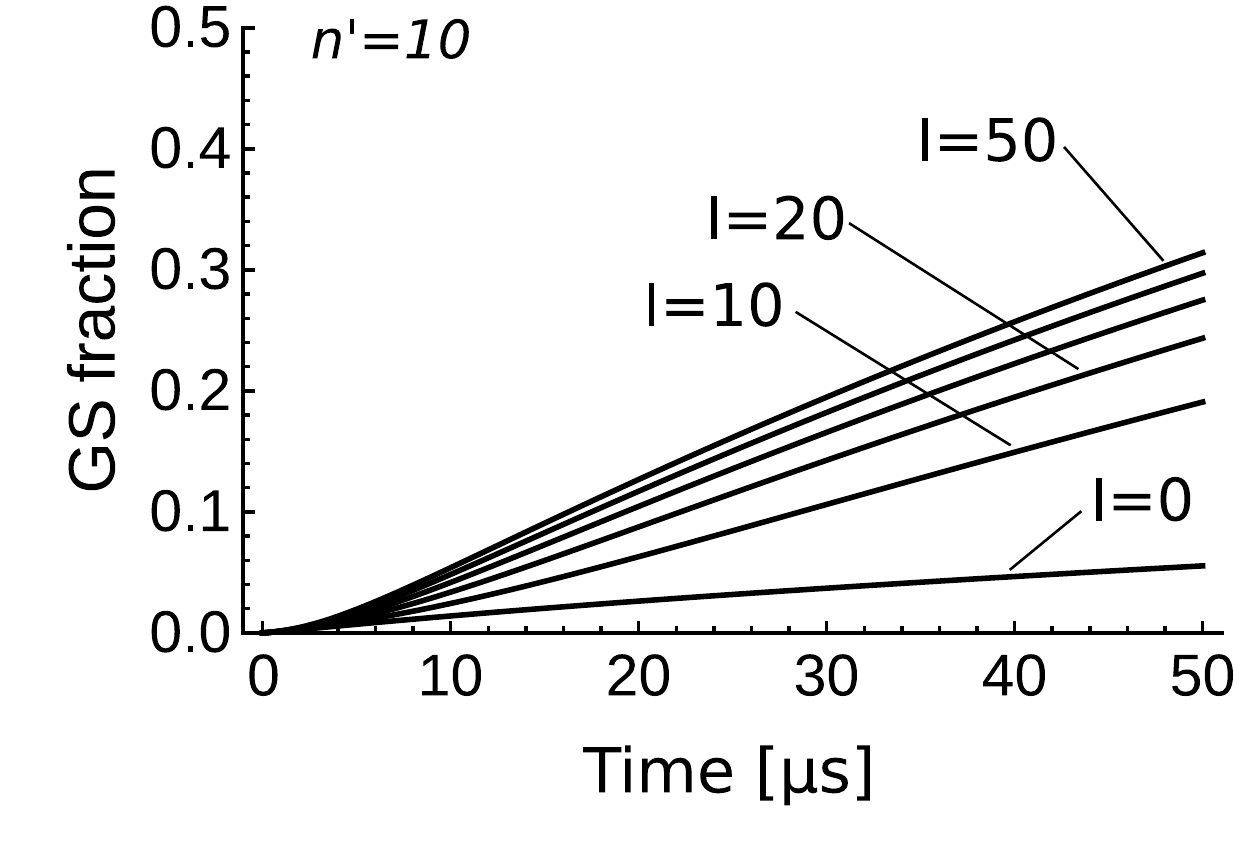}}

\caption{Ground state (GS) fraction as a function of time for THz stimulated decay toward different $n'$-manifold and with different light intensity scalings.
The population is initiated with a statistical distribution of ($m,k$) states in the $n=30$ manifold. 
Unpolarized THz light, with linewidths covering the level broadening given in Fig.~\ref{fig:2a}, stimulates $\Delta n=\pm 1$ transitions down to $n'=5$ (left column) and $n'=10$ (right column).
We implement in the first line a \emph{flat} intensity scaling, i.e. every $\Delta n=-1$ transition is driven with the same intensity. The second (respectively third) line shows the results for a linear \emph{increase} (respectively  \emph{decrease}) scaling resulting in a 100 times more light intensity at the high (respectively low) $n$ transitions. The light intensities ($I$ in units of $\unit[]{W/m^2}$) indicated in the graphs are the total intensities over all stimulated $\Delta n = - 1$ transitions.}
\label{fig:THzMixingPlots}
\end{figure}

The motivation behind the choice of those different scalings lies in the previous observation that a steady state where all involved rates $\Gamma$ are equal is desirable.   
From 
\begin{equation}
\label{eq:intensity_dipole}
    \Gamma \propto \frac{I d_{\rm{eff}}^2}{\Gamma_{\rm L}},
\end{equation}
where $\Gamma_{\rm L}$ is the spectral bandwidth of the light, we see that the light intensity $I$ to drive a $n\rightarrow n-1$ transition scales with the inverse of the square of the effective transition dipole, $d_{\rm{eff}}^{2}$, and is proportional to the bandwidth of the light source. The effective dipole reflects the behavior of the sum of the many dipoles between different states of the $n$ and $n'=n-1$ manifolds. $d_{\rm{eff}}^2$ can be estimated as the average of the squared dipoles between all sub-states of the $n$ and $n-1$ manifold
$d_{\rm{eff}}^2= \frac{1}{n^2} \sum_{k=0}^{n-1} \sum_{m=-k}^{k} d_{n,k,m \rightarrow n-1,k',m'}^2$. The choice of $d_{\rm{eff}}$ is not unique, but in all cases we found that $d_{\rm{eff}}^2$ scales roughly as $n^4$.
Fig.~\ref{fig:2a} indicates that $\Gamma_{\rm L}$ scales as $n^4$ as well. 
Thus, a \emph{flat} scaling of the power appears to be a reasonable choice. The \emph{linear increase} scaling will provide more power toward low $n$ transitions which are harder to drive while the \emph{linear decrease} scaling will provide less power to those transitions, but significantly more at high $n$ which should result in a good and fast mixing of the high ($n,m$) states which are the longest lived ones. Fig.~\ref{fig:THzMixingPlots} shows the obtained ground state fraction as a function of time for different $n'$ values and the three different THz intensity scalings.

Stimulating transitions down to lower $n'$ results in a faster deexcitation, as shown already by Eq.~\ref{eq: DeexcitationScal}. After $\unit[50]{\mu s}$ the ground state population is by a factor $2$ larger for the $n'=5$ than for to the $n'=10$ case. However, driving more transitions requires a higher total light intensity, which is in this case an order of magnitude higher for $n'=5$ than for $n'=10$.

Comparing Fig.~\ref{fig:THzPlot1} and \ref{fig:THzPlot2} shows that the \emph{flat} and \emph{linear increase} scalings provide sensibly the same results. However, the \emph{linear decrease} scaling, Fig.~\ref{fig:THzPlot3}, is significantly worse which indicates that mixing the states in the high $n$ manifolds toward low $m$ states which exhibit fast spontaneous rates is not sufficient to obtain good results if the low $n$ transitions are not driven fast enough. Additionally, higher power at high $n$ also leads to a decreased performance due to the enhanced excitation mechanism to manifolds with $n>30$.

The simulations showed that close to $\unit[80]{\%}$ of the atoms initially distributed in the $n=30$ can be brought to ground state within $\unit[50]{\mu s}$ with total intensities of the order of $\sim \unit[200]{W/m^2}$. However, since the proposed mechanism couples all states between $n=35$ and $n'=5$, the atoms initially populating those states will also be deexcited leading to an even higher total number of atoms in ground state. The exact fraction achievable and the possible power scaling optimizations depend on the particular initial state distribution, but given the versatility of the proposed technique, the targeted parameters can easily be adapted to different experimental conditions.

To achieve the best efficiency in the presently studied case, one needs to drive 25 (from $n=30$ to $n'=5$) sharp transitions which is not straight-forward to realize experimentally in particular with the total output power indicated by the simulation. Several options are nevertheless possible. For the application foreseen we found that a photomixer, which converts the output of two continuous-wave (cw) light sources with adjacent frequencies $\nu_{\rm i}$ and $\nu_{\rm j}$ into cw THz radiation at exactly the difference frequency $\nu_{\rm i} - \nu_{\rm j}$, seems to be a good choice to generate that many frequencies \cite{2011JAP...109f1301P} especially since the multi-frequency input laser light can be produced using pulse shaping from a single broadband laser source \cite{liu1996terahertz,metcalf2013fully,hamamda2015ro,finneran2015decade}. We have hence investigated what results can be obtained by an off-the-shelf photomixer which was already tested on a beam of Rydberg caesium and successfully demonstrated the stimulated deexcitation of the atoms \cite{vieil2018, vieil2020}. The output power of such devices is in the $\unit[]{mW}$ range at $\sim\unit[200]{GHz}$, but drastically decreases toward the higher frequency region \cite{Latzel2017} such that it becomes unfit to the purpose for $n'<15$.
However, photomixers can be an attractive solution in cases where $n'$ can be chosen relatively high as in beam experiments where a long flight path separates the formation from the measurement region. In that case it can be shown that with typical powers the ground state fraction can be improved right after formation by a factor $\sim 2$. Since it additionally spreads the initial distribution in $n=30$ toward lower lying $n'$ manifolds within a few $\unit[]{\mu s}$ it results in a very significant gain of ground state atoms after spontaneous decay throughout the flight time toward the measurement region.

\subsection{\label{s:RF} Microwave stimulated mixing}
We saw that the \emph{flat} scaling that homogeneously distributes the power among the $\Delta n = \pm1$ transitions is relatively efficient, but requires high total THz powers, and that a \emph{linear decrease} scaling, which could potentially be more efficient in mixing the $m$ states of the high $n$-manifolds, leads instead to losses through excitation. A potential improvement could hence be to reduce the THz power to a level that might still be enough to deexcite the population, but insufficient to mix the intra-manifold levels, and to add microwave radiation to mix the ($m,k$) levels instead. This would have the advantage to avoid ionization and excitation and be experimentally easy to implement since $\Delta n = 0$ transition dipoles are large and powerful microwave sources exist.

The optimal microwave power scaling to efficiently transfer the population between the energy levels of a given $n$-manifold is complex to determine (see all transitions in Fig.~\ref{fig:cusp}). Thus, we performed the simulations by implementing single unpolarized broadband sources with a large bandwidth covering the entire $n$-manifold. 
The simulations however showed that the addition of microwaves was only marginally increasing the ground state fraction (sub $5\%$ level) highlighting that the bottleneck, in this scheme, resides in accessing low $n$-states as was already suggested by the comparisons of $n'=5$ and $n'=10$ and the \emph{flat} and \emph{linear decrease} scalings in Fig.~\ref{fig:THzMixingPlots}.
For particular experimental conditions the optimal choice might still be the use of microwave light. In the case of CE production for example, where the principal quantum number distribution can be small and well controlled, a scheme relying on the sole use of microwaves to mix the angular momenta coupled to a laser (cf.~section \ref{s : laser}) can be a very promising choice. 

\subsection{\label{s : laser}Laser stimulated deexcitation}	
The coupling of a large number of states with microwave and THz light to efficiently mix and deexcite an initial population of Rydberg antihydrogen atoms, as studied in section \ref{s:THz} and \ref{s:RF}, is particularly interesting as it can address a large distribution of states up to $n\sim35$. However, to be fast and efficient ($<\unit[50]{\mu s}$), Eq.~\ref{eq: DeexcitationScal} indicates that low-lying $n'$ states, that rapidly decay spontaneously, need to be reached which requires the generation of a large number of frequencies (typically $\sim 20$) in the range of a few \unit[]{mW} per transition which remains experimentally challenging at frequencies $>\unit[1]{THz}$ \cite{Latzel2017}.

An alternative can be the coupling of the aforementioned scheme, restricted to a few initially populated levels, to a laser which can drive the mixed population, for instance, toward the $n''=3$ manifold where levels have a spontaneous lifetime of the order of \unit[10]{ns}. The $2p$ state with a lifetime of \unit[1.6]{ns} may as well be targeted and would lead in theory to better results, but large power at this UV wavelength ($\unit[368]{nm}$ from $n'=20$) is much more challenging to reach than for the $\unit[840]{nm}$ wavelength from $n'=20$ to $n''=3$. 

\begin{figure}
\centering
\subfloat[\label{fig:LaserPlot1}]{\includegraphics[width=0.5\linewidth]{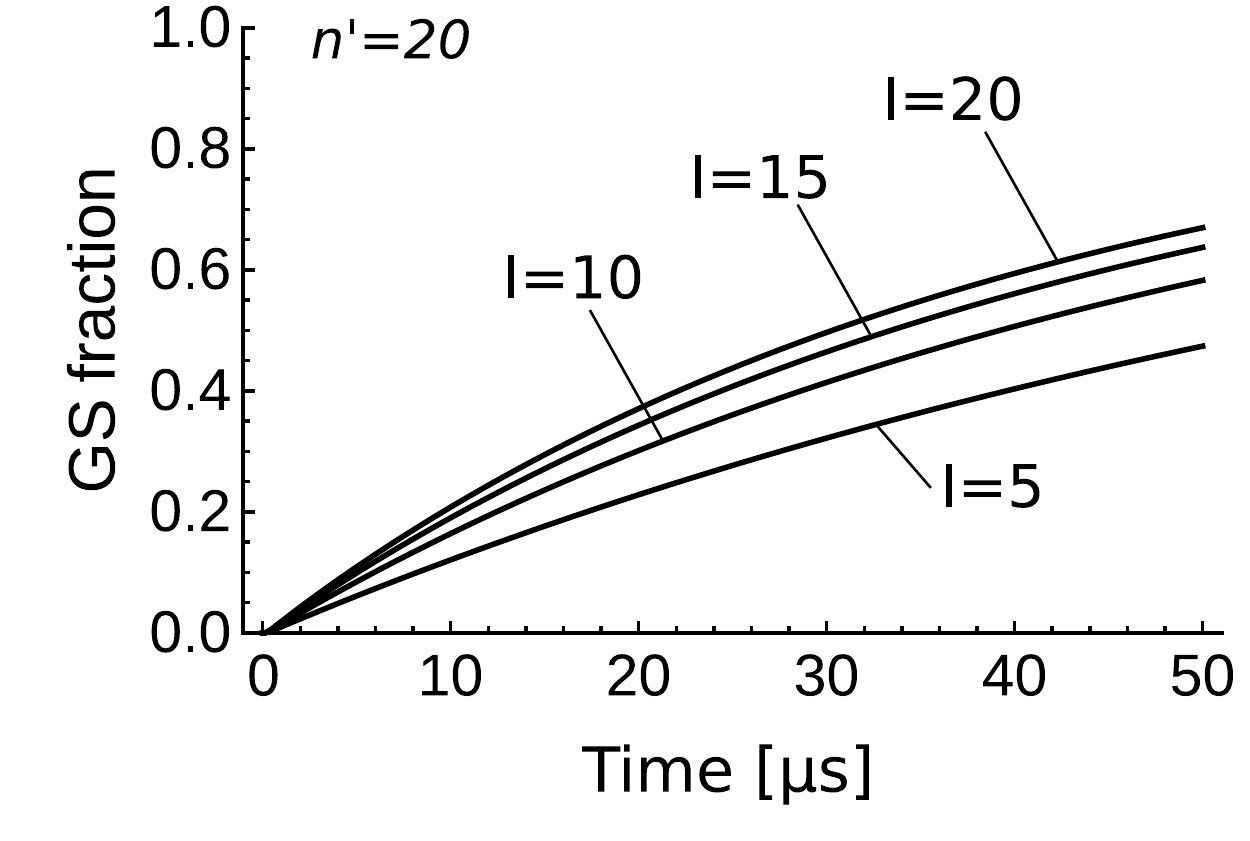}}
\subfloat[\label{fig:LaserPlot2}]{\includegraphics[width=0.5\linewidth]{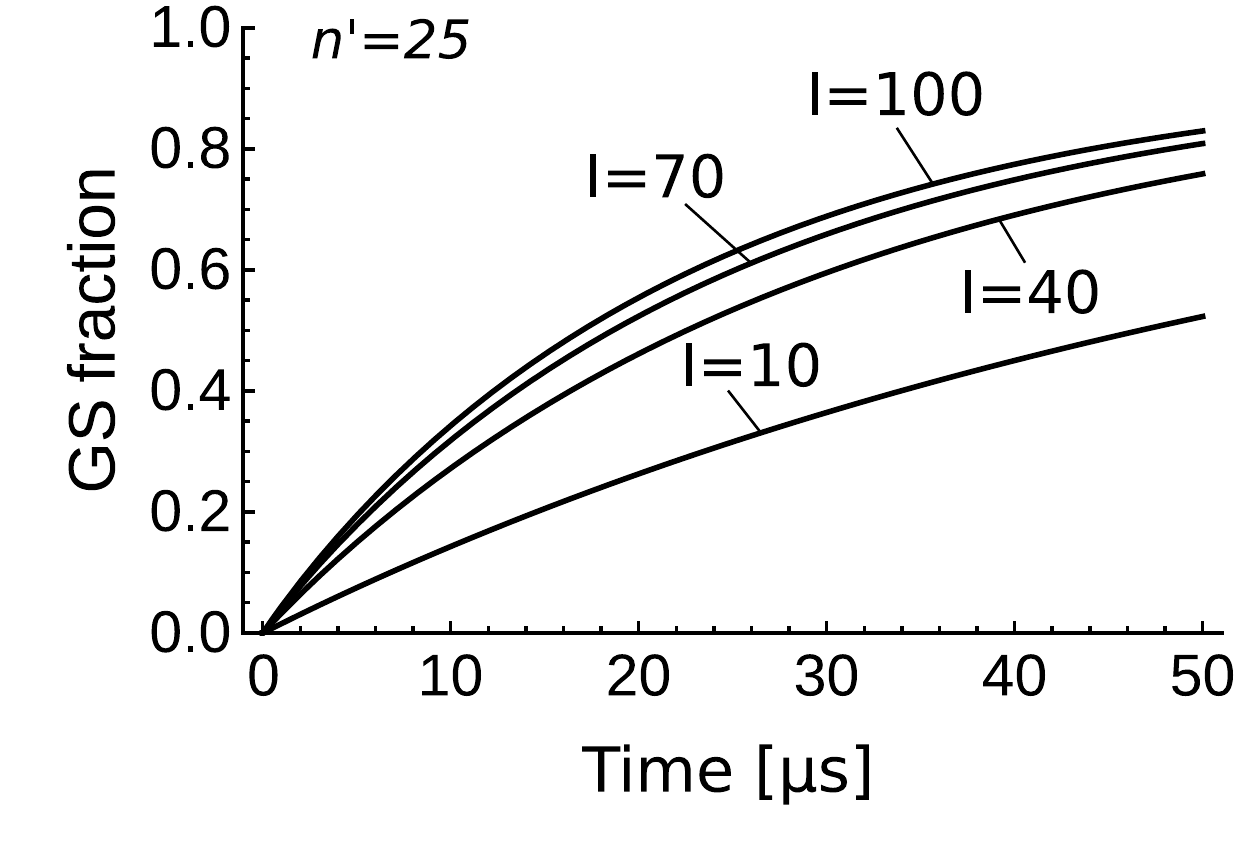}}

\subfloat[\label{fig:LaserPlot3}]{\includegraphics[width=0.5\linewidth]{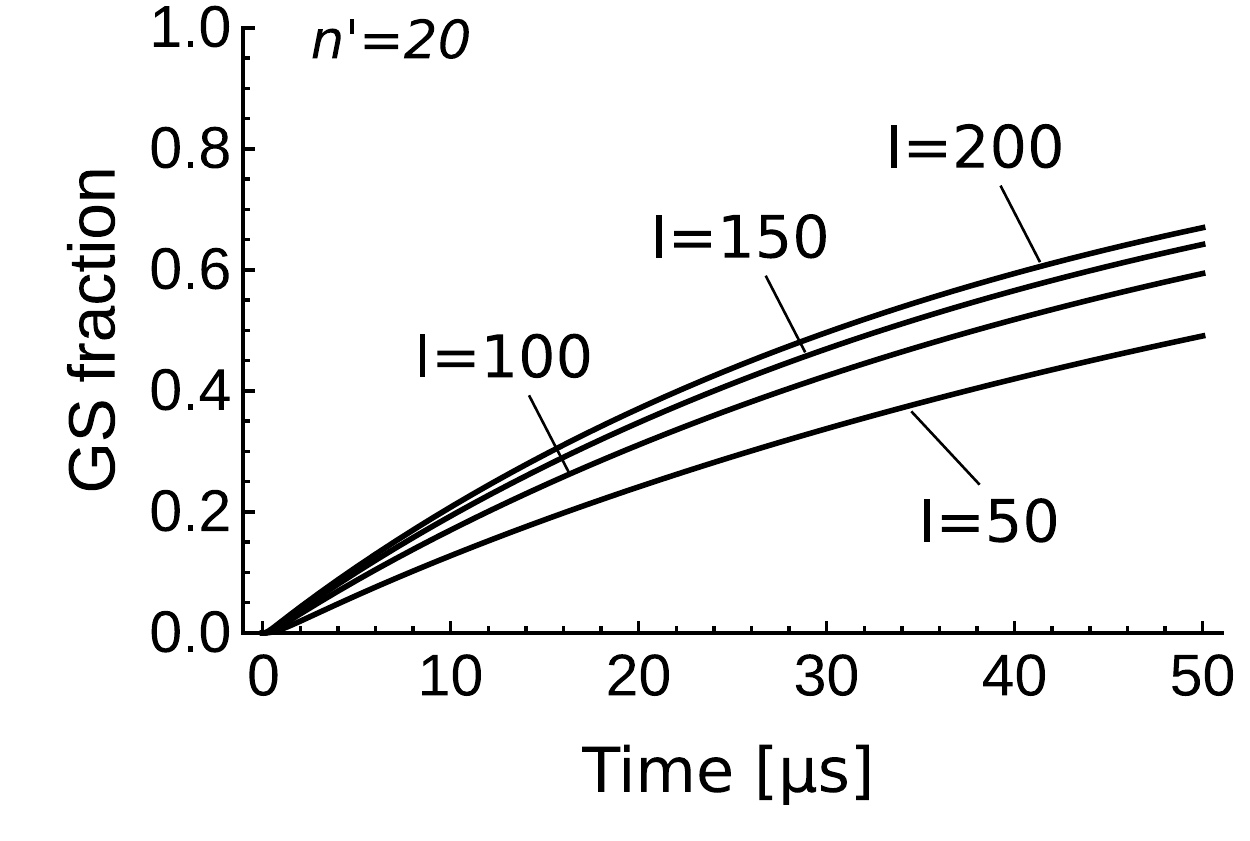}}
\subfloat[\label{fig:LaserPlot4}]{\includegraphics[width=0.5\linewidth]{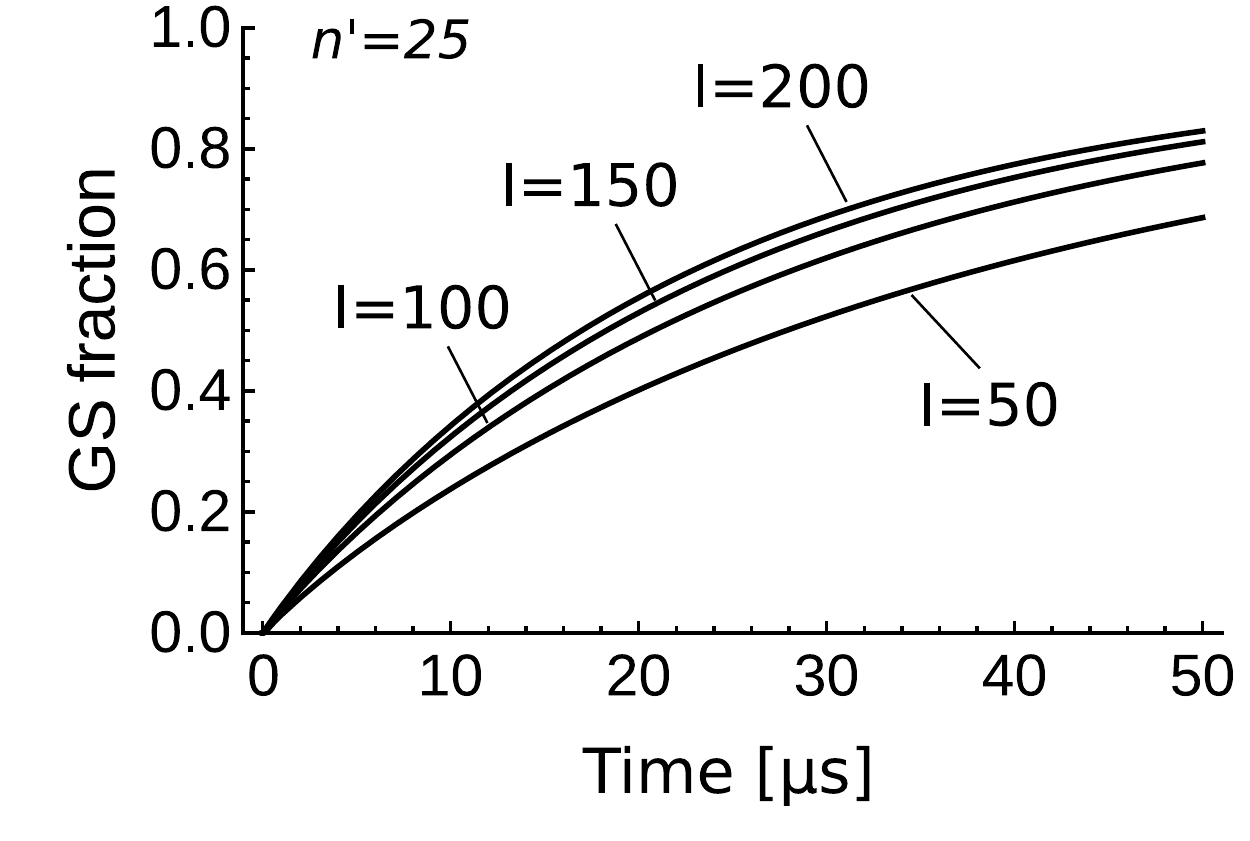}}

\subfloat[\label{fig:LaserPlot5}]{\includegraphics[width=0.5\linewidth]{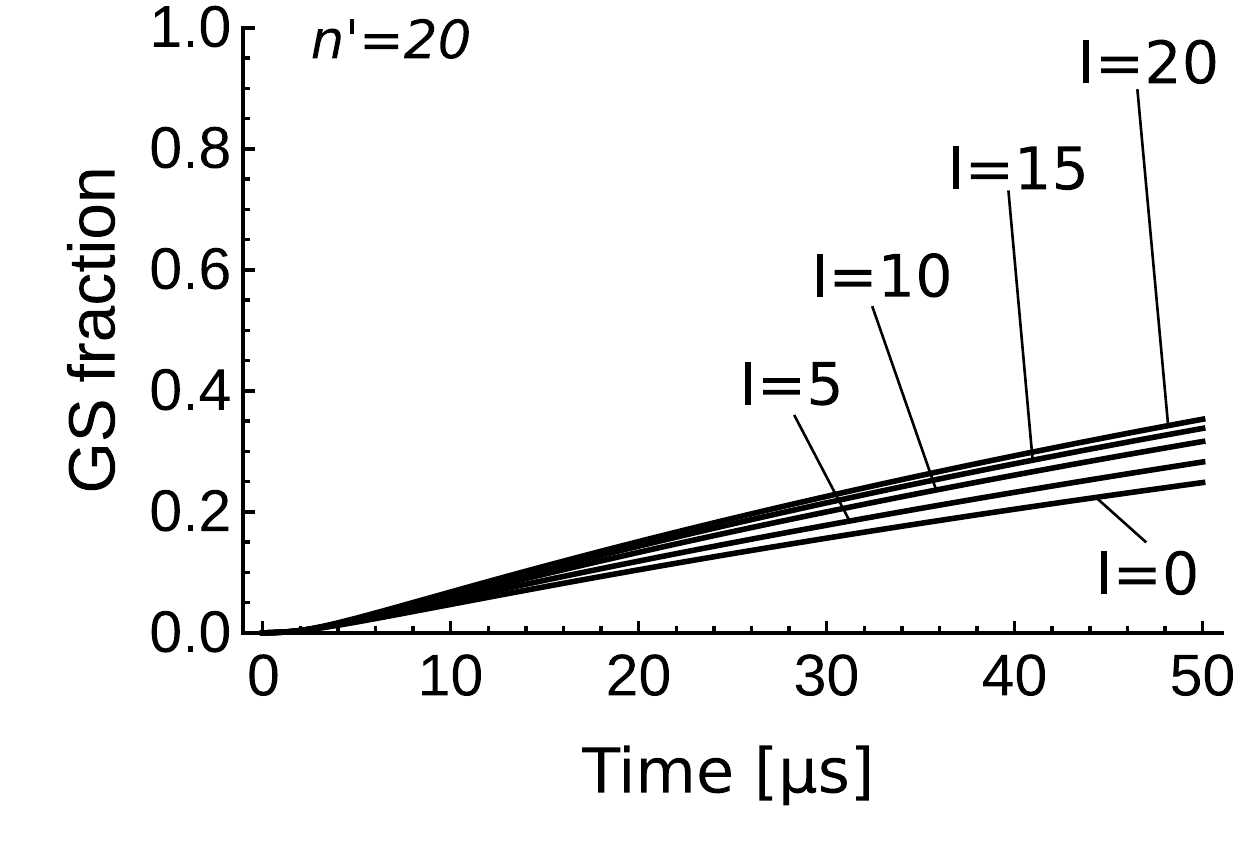}}
\subfloat[\label{fig:LaserPlot6}]{\includegraphics[width=0.5\linewidth]{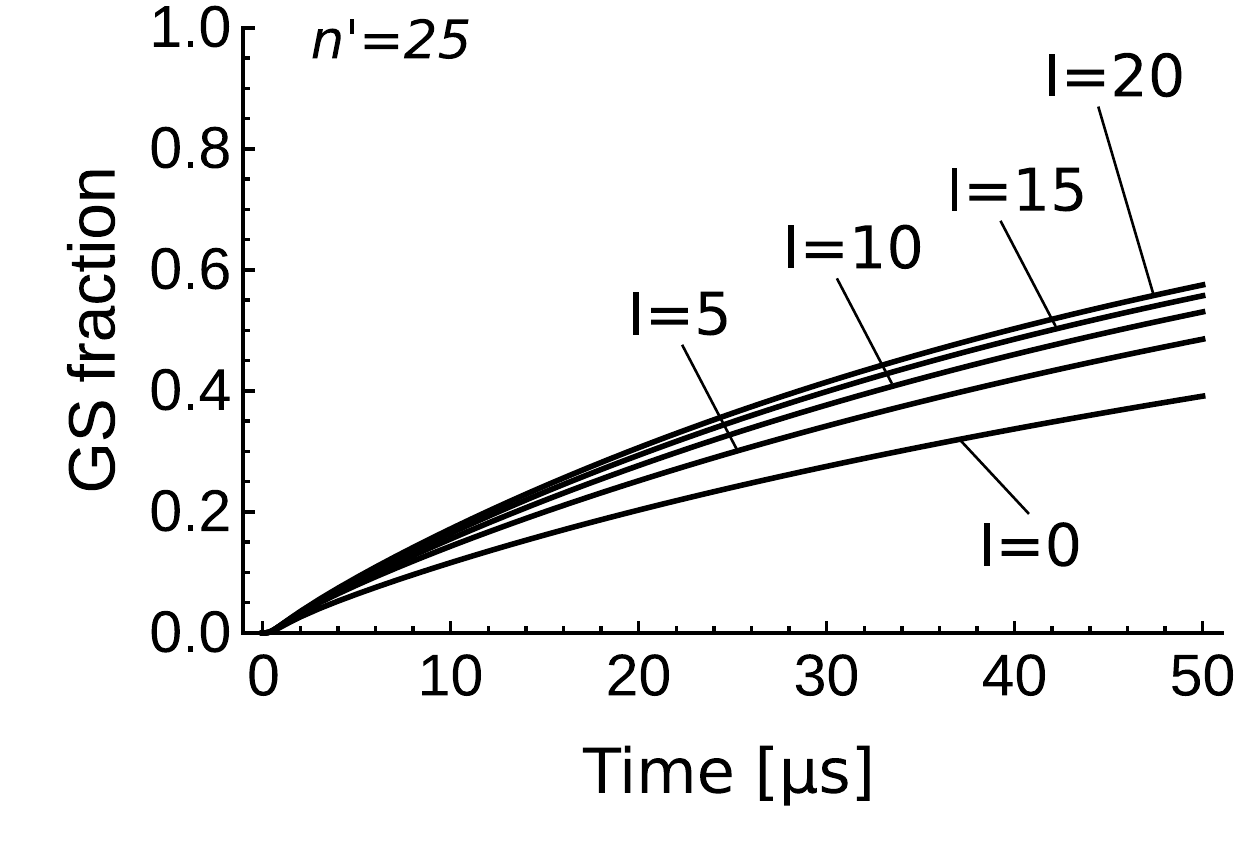}}
\caption{Ground state (GS) fraction as a function of time for different scenarios - using THz and laser (first two lines) or THz, laser and microwaves (third line). The population is initiated in $n=30$. Transitions are stimulated toward $n'=20$ (left column) and $n'=25$ (right column). A laser couples the $n'$ manifold to $n''=3$. First line: GS fraction for different laser intensities and a given total THz intensity of $\unit[200]{W/m^2}$. The laser intensities are given in $\unit[10^5]{W/m^2}$ per $\unit[500]{MHz}$ laser bandwidth $\Gamma_{\rm{L}}$. Second line: GS fraction for different total THz intensities and a fixed laser intensity of $\unit[20 \times 10^5]{W/m^2}$ (left) and $\unit[100 \times 10^5]{W/m^2}$ (right). Third line: GS fraction for a reduced THz intensity of $\unit[10]{W/m^2}$, a fixed laser intensity of $\unit[20 \times 10^5]{W/m^2}$ (left) and $\unit[100 \times 10^5]{W/m^2}$ (right) and different total microwave
intensities.}
\label{fig:LaserPlot}
\end{figure}

We simulated this scheme by solving, like before, the rate equations for all $(n,k,m)$ levels up to $n = 35$ in a magnetic field of $\unit[1]{T}$. The population is initiated in the $n=30$ manifold. As stressed before, the experimental distribution of states is not precisely known. Therefore, in order to take into account several possible initial distributions, we present results for $n'=20$ (left column) and $n'=25$ (right column) in Fig.~\ref{fig:LaserPlot}.

In the first line, we investigate the required laser intensities when equipopulating the $n$ to $n'$ levels with a THz intensity of $\unit[200]{W/m^2}$ which is rather high, but originates from the necessary repopulation of the $n'$ levels that are coupled to the laser at a rate which is of the same order as the laser deexcitation rates. We treat the case of a broadband laser where all sublevels of the $n''=3$ manifold are coupled to the Rydberg state distribution. With a choice of $\unit[500]{MHz}$ (FWHM for a Lorentzian spectrum) all states in the $n''=3$-manifold are coupled to at least one level in the $n'=20$ or $n'=25$-manifold. Since the transitions in question have similar dipoles and the $n'$-manifold is constantly equipopulated, the plot is generic and other linewidth choices will simply scale the power. The laser intensities indicated are in the range of $\unit[]{kW}$ for a typical $\unit[]{cm^2}$ spot and could be produced by cavity-enhanced or pulsed lasers. Photoionization at those wavelengths is small \cite{COM181}.

The second line in Fig.~\ref{fig:LaserPlot} shows the obtained ground state fraction for a given laser and different THz intensities. The previously chosen $\unit[200]{W/m^2}$ is close to the saturation limit. However, lowering the THz intensity rather quickly decreases the performance.

Consequently, we show in the third line how microwaves can compensate for the losses introduced by a reduced THz intensity. Indeed, once a laser drives the population to low $n''$, the bottleneck does not anymore reside in the deexcitation via THz, but rather in the fast repopulation of the states depopulated by the laser; in that case the additional use of microwaves can be useful. We see that $60\%$ of the atoms can be brought to ground state with a much lower THz intensity of $\unit[10]{W/m^2}$ when adding microwave radiation. The required microwave power is comparatively small (total intensities of the order of $\unit[20]{W/m^2}$) and can be easily available in an experiment. For simplicity, we implemented a \emph{flat} microwave scaling.

We conclude that for the given cases close to unity ground state fractions can be reached even faster than in the previous scheme if the THz transitions are limited to only five frequencies at the expense of a larger laser intensity driving, in that case, the $n'=25\rightarrow n''=3$ transitions. With this particular implementation an initial distribution of states between $n=35$ and $n=25$ can be addressed. Lowering the THz power can be to some extend compensated by adding microwaves. However, the inter-manifold mixing remains crucial.

\section{Stimulated radiative recombination within the positron-antiproton plasma}
\label{s:radrecomb}

\subsection{Collisional mixing}\label{s:collis}

In the previous sections we have assumed that the antihydrogen atoms formed have moved out of the plasma. However, within typically \unit[$\sim$1]{$\mu$s} (plasma radius of \unit[$\sim$1]{mm} and antihydrogen velocities of \unit[$\sim$1000]{m/s}), the formed atoms are still contained within the plasma where they encounter collisions with positrons and antiprotons. Collisions have been suggested to be efficient in mixing the $(n,k,m)$ states \cite{ROB08,radics2014scaling,gerber2018modeling,vrinceanu2017treatment}.
The antihydrogen atoms in the plasma are exposed to a Stark effect produced
by the electric field $\bm E$ which is created by the other particles present (positrons, antiprotons). We will see below that the dominant effect is produced by the collisions with the fast positrons. The electric field
is given by $\bm E\sim e^2/4\pi \varepsilon_0 R^2$ where $R$ is the distance between the charged particles and the Rydberg antihydrogen atom.
State changing arises when the Stark shift ($\sim  n^2 \bm E$ in atomic units) reaches the energy separation between the initial and final states having different $n$, $k$ or $m$ quantum numbers.
Because, for $n\sim30$ and below, the $n$ levels are well separated in energy, the $n$ changing collisions require a rather large electric field to be efficient. Thus, we can consider the produced Stark effect to be dominant over the Zeeman effect. Hence, the $n$ mixing rate will be well approximated by the magnetic field-free case, so with a
collisional rate for the $n\rightarrow n'$ deexcitation of the order of $\unit[10^{-6}]{cm^3/s} \left(\frac{T_{e^+}}{\unit[10]{K}}\right)^{-0.17} {n'}^{6.66}/{n}^5$   \cite{mansbach1969monte,devos1979transitions,vriens1980cross,pohl2008rydberg}. Therefore, for a $\unit[10]{K}$ positron plasma at a density of $\unit[10^8]{cm^{-3}}$, and $n\sim n' \sim 30$, the $n$-mixing will happen on a timescale of several tens of $\unit[]{\mu s}$ and so will be negligible within the $\unit[1]{\mu s}$ timescale of the presence of the antihydrogen within the plasma. The mixing achieved is of course dependent on the details of the plasma and we note that other parameters (for example long and dense plasmas) may enhance the effect \cite{radics2014scaling,radics2016antihydrogen}.

Mixing $m$ and $k$ levels requires much smaller fields than for the $n$ manifolds, thus the electric field present in the plasma can lead to a sensible effect \cite{gallagher1994}. The energy separation is of the order of $\unit[14]{GHz}$ (at \unit[1]{T}) for $m$ states and only of the order of \unit[100]{MHz} for $k$ states, as indicated by Fig.~\ref{fig:cusp}. Using the relations above, we thus find that $k$-mixing requires an impact parameter of $
R\sim \unit[13]{\mu m}$ that is also, fortuitously, the value for the  typical (Wigner-Seitz) inter-particle distance $\left({\frac {3}{4\pi n_{e^+}}}\right)^{1/3}$.  For the $m$-mixing, we find $R\sim \unit[1]{\mu m}$ that is incidentally very close to the value for the classical distance of closest approach in positron-positron Coulomb scattering $\frac{e^2}{4 \pi \varepsilon_0 k_B T}$.
In all cases, the mixing collisions arise with impact parameters $R$ much larger than the antihydrogen Rydberg size  $\sim n^2 \times \unit[0.05]{nm}$  which justifies our simple treatment of the Stark effect.

The efficiency of the level crossing is harder to estimate. However, a simple Landau-Zener model indicates that the level crossing should be efficient because the collisional time $R/v$,
estimated using the typical $v=\sqrt{k_B T/m} \sim  \unit[10^4]{m/s}$ positron velocity, is comparable with the frequency spacing at crossing that is the Rabi frequency  $ \int \langle nkm| e \bm r . \bm E(R(t)) |nk' m' \rangle/\hbar$.
In conclusion, a reasonable estimate of the collisional mixing rate is
 $n_{e^+} \pi R^2 v$.

Therefore, within the plasma, we find collisional rates of the order of one per several tens of microseconds for $n$-mixing,
of a few per microsecond for $m$-mixing and almost one per nanosecond for $k$-mixing. 
During the microsecond traveling time of the antihydrogen within the plasma, we will thus, for simplicity, neglect $n$-mixing as well as $m$-mixing (since it will, at best, only be partial), but will assume a complete mixing of the $k$ states.

\subsection{Stimulated radiative recombination}

In the previous sections we have investigated how to couple a large number of levels to other ones which rapidly spontaneously decay to ground-state. Within the plasma,
the case of stimulated radiative recombination (srr) where a laser drives a positron 
from the continuum directly to a bound state, is a very similar process, if we treat the many levels mentioned before as the continuum. Stimulated radiative recombination has already been proposed as an efficient way to form antihydrogen \cite{budker1978electron,neumann1983laser,GABRIELSE1988,wolf1993laser,muller1997production}. Following the realization on hydrogen \cite{yousif1991experimental,articleRoge}, a stimulated formation of antihydrogen has been attempted using a CO$_2$ laser down to $n'=11$, but without success \cite{amoretti2006search}. The invoked explanation involved the competition with the   three body recombination (3BR)  that populates several (mostly very excited) levels at a rate \cite{ROB08}
\begin{equation}
\Gamma^{\rm 3BR} \sim \unit[160]{s^{-1}}  \left(\frac{n_{e^+}}{ \unit[10^8]{cm}^3} \right)^2 \left( \frac{\unit[10]{K}}{T} \right)^{4.5} \label{3BR_rate}
\end{equation}
for a non-correlated plasma \cite{2000PhLA..264..465H}. We note that experimental rates have been measured to be different under specific plasma conditions \cite{Fujiwara2008}. 
\subsubsection{Theory}
\begin{figure}
\centering
\includegraphics[width=1\linewidth]{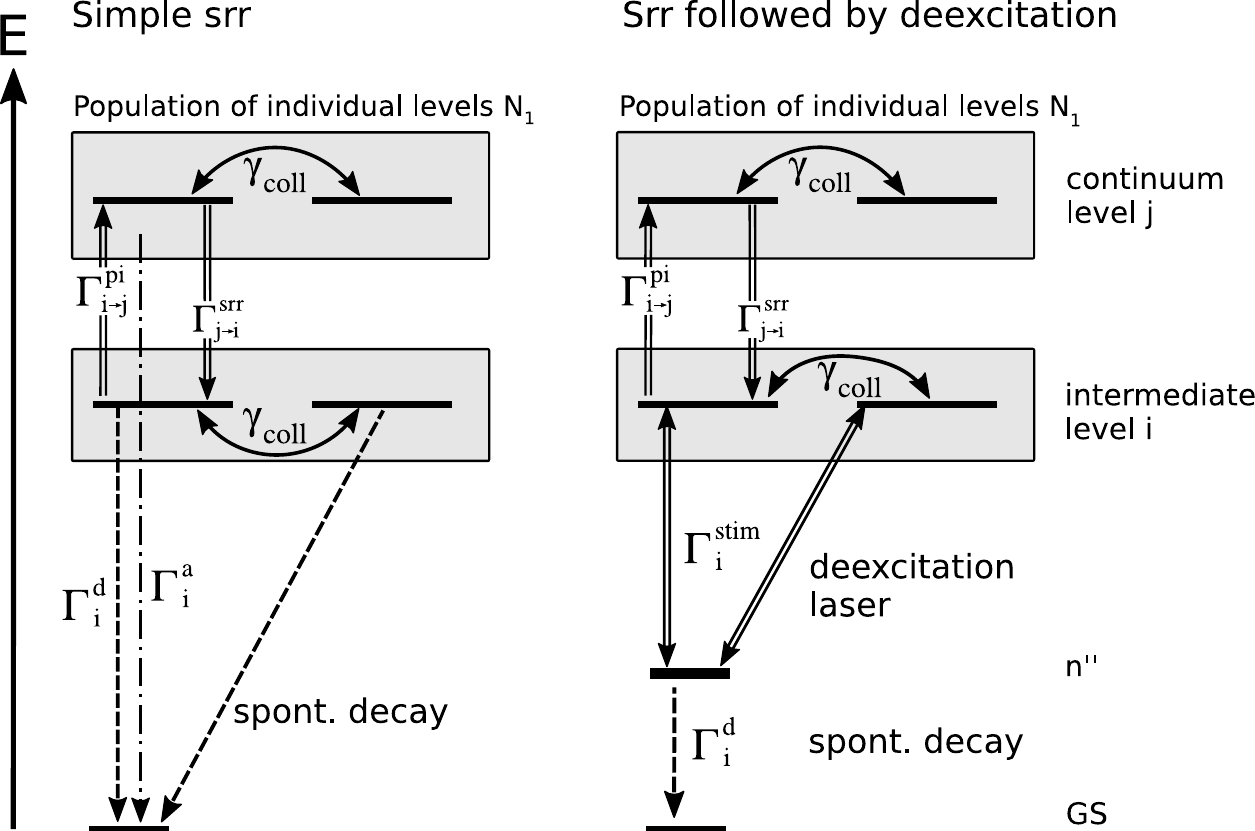}
\caption{Illustration of the nomenclature used for the decay rates involved in the ``simple'' srr process (left) or srr followed by deexcitation (right)  including:  photoionizing (pi), decay (d), association (a) and stimulated (stim) rates. The decay rate can be due to spontaneous emission, collision, ejection out of the laser zone, etc. The collisional rates responsible for the $k$-mixing in the bound states and population reshuffling  in the continuum are indicated by $\gamma_{\rm coll}$.}
\label{fig:srr_rates} 
\end{figure}

In the stimulated radiative recombination case, a laser of frequency $\nu$, irradiance $I = \int I(\nu) {\rm d} \nu $ and a waist of $\unit[1]{mm}$, to cover the plasma, 
couples bound states to continuum states with a (positron) energy above ionization threshold: 
$E=  \kappa^2 R_y = h  \nu - R_y/n'^2 = \frac{1}{2} m v^2$.
As studied in more detail in the appendix section \ref{s:appendixssrr}, the standard srr theory, a theory based on the photoassociation analogy, or a simple rate equation model illustrated in Fig.~\ref{fig:srr_rates}, lead to the same results. The rate equation model is the more general and simpler one which is why it will be used hereafter.
When driving several levels j toward a level i, the
srr rate  $\Gamma^{\rm{srr}}_{\rm{i}}  =  \sum_{\rm{j}} \Gamma^{\rm{srr}}_{\rm{j}\rightarrow\rm{i}} $ competes with 
the photoionization rate $\Gamma^{\rm{pi}}_{\rm{i}} =\sum_{\rm{j}} \Gamma^{\rm{pi}}_{\rm{i}\rightarrow\rm{j}}$ and the decay rate  $\Gamma^{\rm{d}}_{\rm{i}}$ from this level. The decay rate can be due to spontaneous emission, collisions, ejection  out of the laser zone, etc.

The srr rate between non degenerate levels is
$\Gamma^{\rm{srr}}_{\rm{j}\rightarrow\rm{i}} = N_1 \Gamma^{\rm{pi}}_{\rm{i}\rightarrow\rm{j}} $,
where $N_1 = n_{e^+} Q_T^{-1}   e^{-E/k_B T}$ is
the population of one level (summing over the  electron spin) in the continuum, 
with  $n_{e^+}$ being the positron plasma density and $Q_T= \left( \frac{2\pi m k_B T}{h^2} \right)^{3/2}$ the translational partition function ($\unit[\sim 7.6\times 10^{16}]{cm^{-3}}$ at \unit[10]{K}).
The phase space density $n_{e^+}Q_T^{-1}$, also called the plasma degeneracy parameter,  can be seen as  the (maximum) population of an individual level in the continuum (the electron spin being summed over). This is the key parameter that ultimately limits the population transfer of continuum states toward the ground state.

Whatever mechanism is used (microwave, THz, laser, collisional mixing etc.) to drive the population to the ground state, the maximum population rate per targeted decayed level will be given by 
$   \Gamma^{\rm{a, max}}_{\rm single\ level} = n_{e^+}Q_T^{-1} \Gamma^{\rm{d}}$.
This optimal association rate toward a single level is reached when the laser is tuned just at resonance ($E=0$ so with $N_1$ maximal).
This is a very similar situation to that of the deexcitation of bound levels, the only difference in the rate equations being that the population of the level is not $1/N$ (one antihydrogen is formed but we do not know in which level), where $N$ is the number of states which are coupled to the short-lived states in the deexcitation case,  but $ n_{e^+}Q_T^{-1} \sim  1.3\times 10^{-9} $ (for numerical values we assume a $\unit[10]{K}$ positron plasma at density $\unit[10^8]{cm^{-3}}$). Therefore, if the association/deexcitation is driven to the same fast-decaying final state, the dynamics of the srr will be $\sim10^{-5}$ times slower than the deexcitation through the coupling of $10000$ states (there are for example $7000$ states between the $n=30$ and $n=20$-manifolds).
As in section \ref{s : laser}, a cascade which ends at $n'=3$ would however lead to a fast $\Gamma^{\rm{d}} \sim \unit[1/10]{ns^{-1}}$. As mentioned before, a stimulated decay down to $n'=2$ would be even faster and would lead to 

 \begin{equation}
    \Gamma^{\rm{a, max}}_{\rm single\ level} = n_{e^+}Q_T^{-1} \Gamma^{\rm{d}} \approx \unit[1]{s^{-1}} \label{max_srr_rate},
\end{equation}
where for numerical values we have used $\Gamma^{\rm{d}}= \unit[1/1.6]{ns^{-1}}$.

\subsubsection{Simple srr}
In the simplest case (as was reported in \cite{amoretti2006search}) a single laser drives a population from the continuum states toward the levels of the $n'$ manifold from which the atoms, by spontaneous decay, eventually reach ground state.

As discussed before, given the diamagnetic spread and collisional mixing, we assume a fast and complete $k$  mixing leading to an average of the srr and spontaneous emission rates in a given ($n',m'$) manifold.

\begin{figure}
\centering
\subfloat[]{\includegraphics[width=0.5\linewidth]{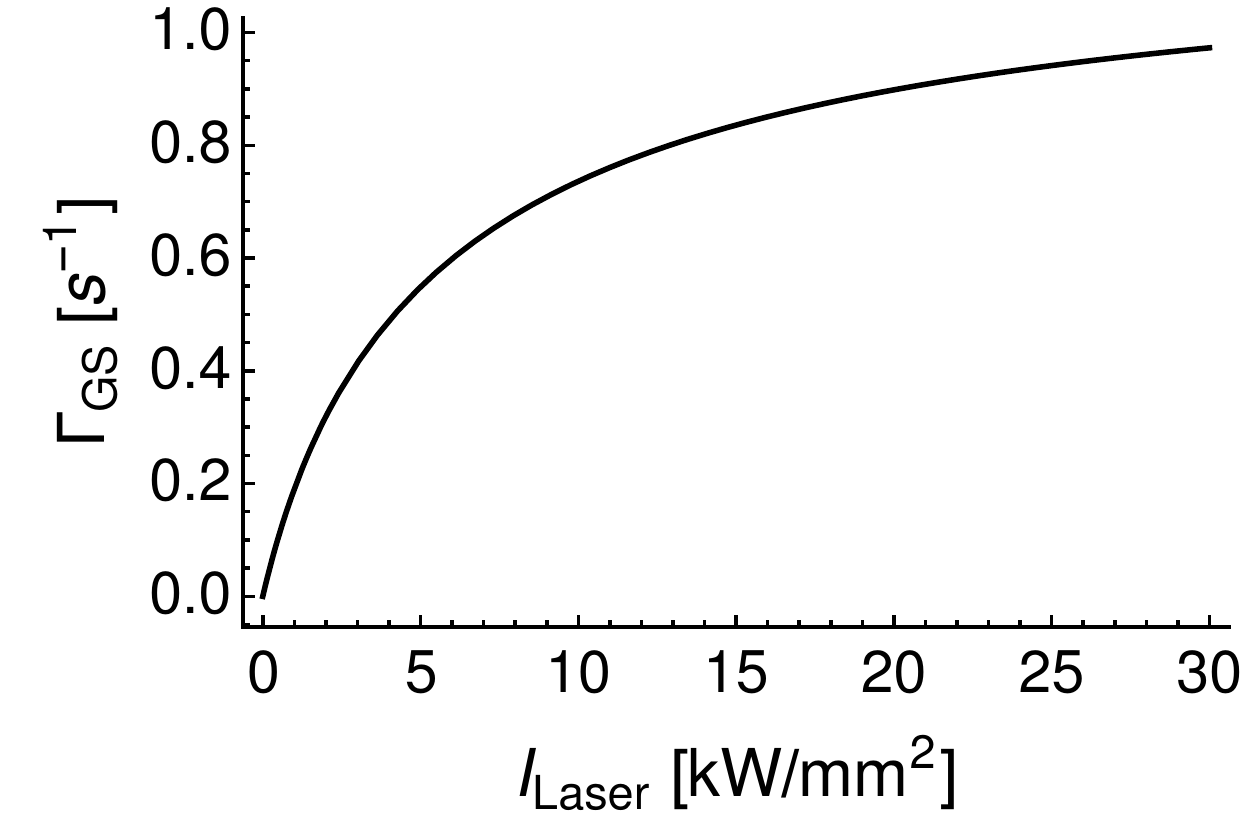}}
\subfloat[]{\includegraphics[width=0.5\linewidth]{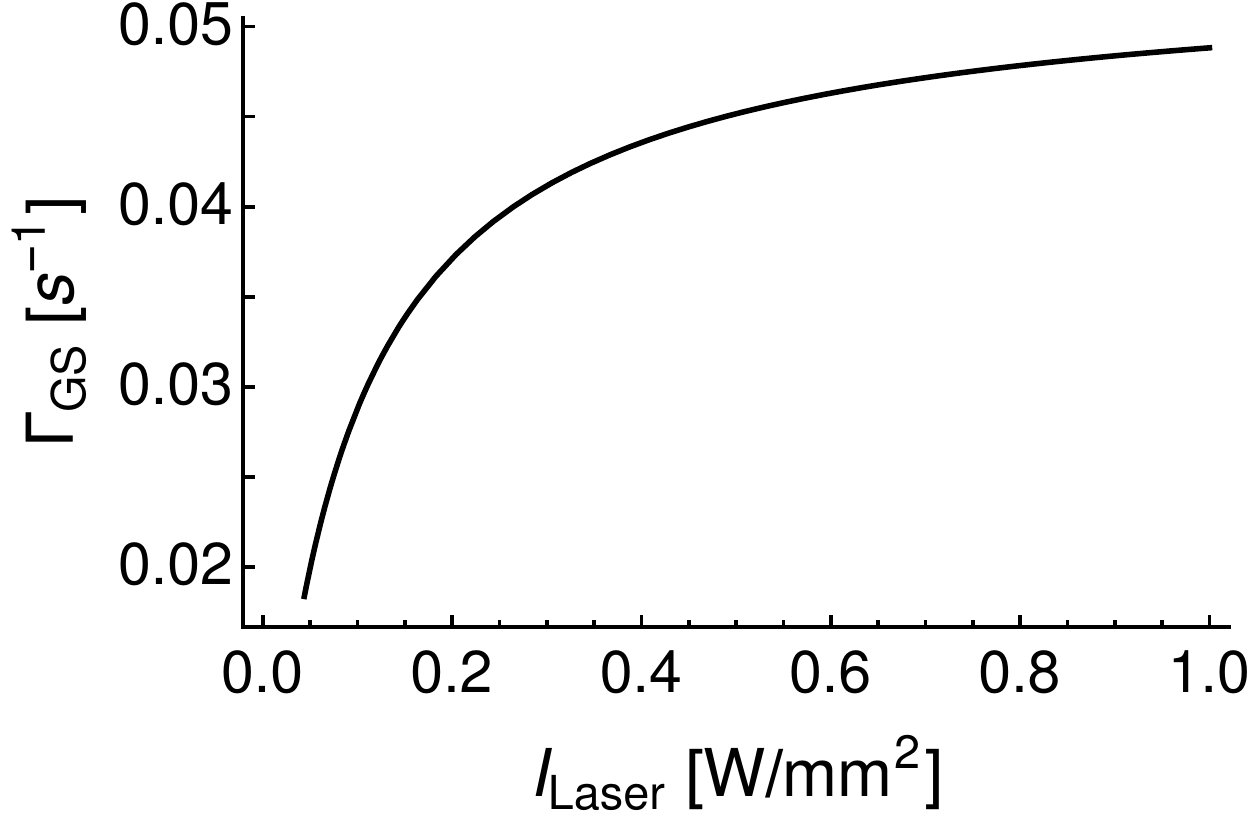}}

\subfloat[]{\includegraphics[width=0.5\linewidth]{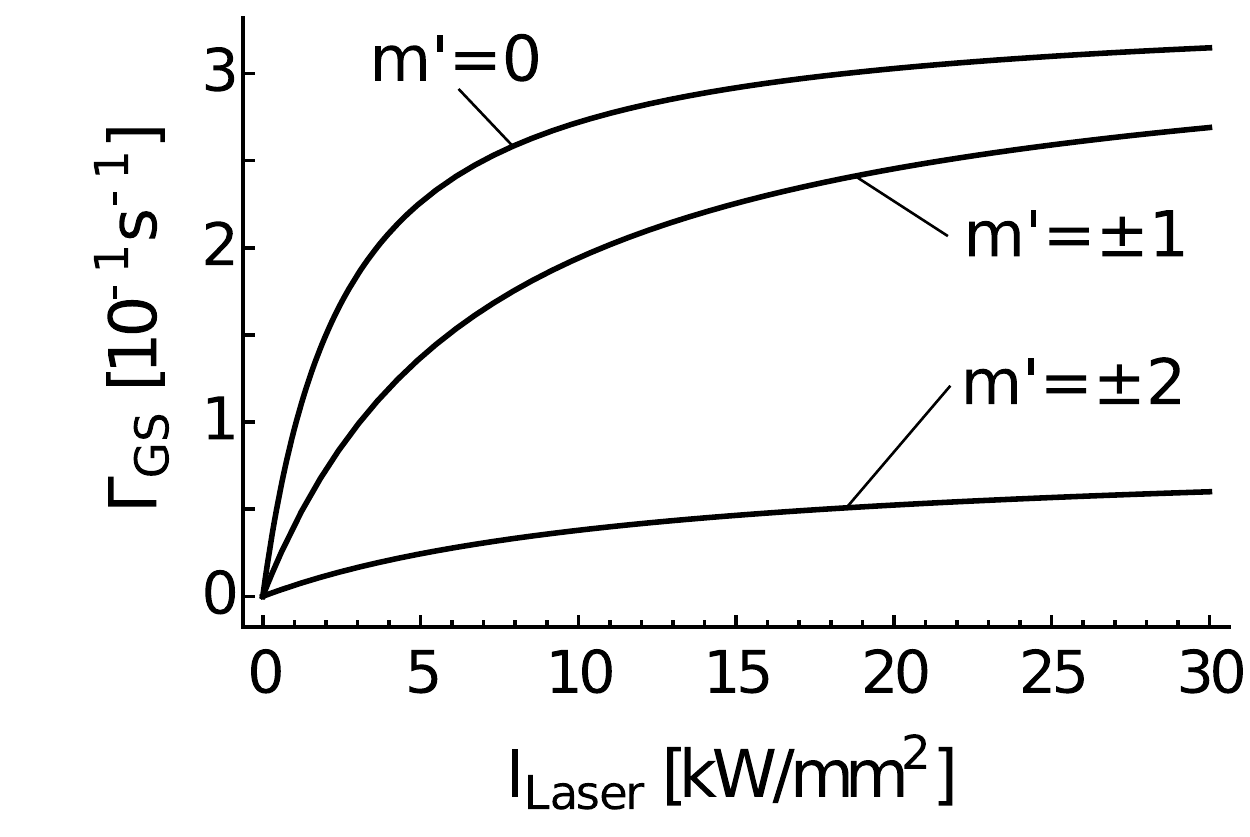}}
\subfloat[]{\includegraphics[width=0.5\linewidth]{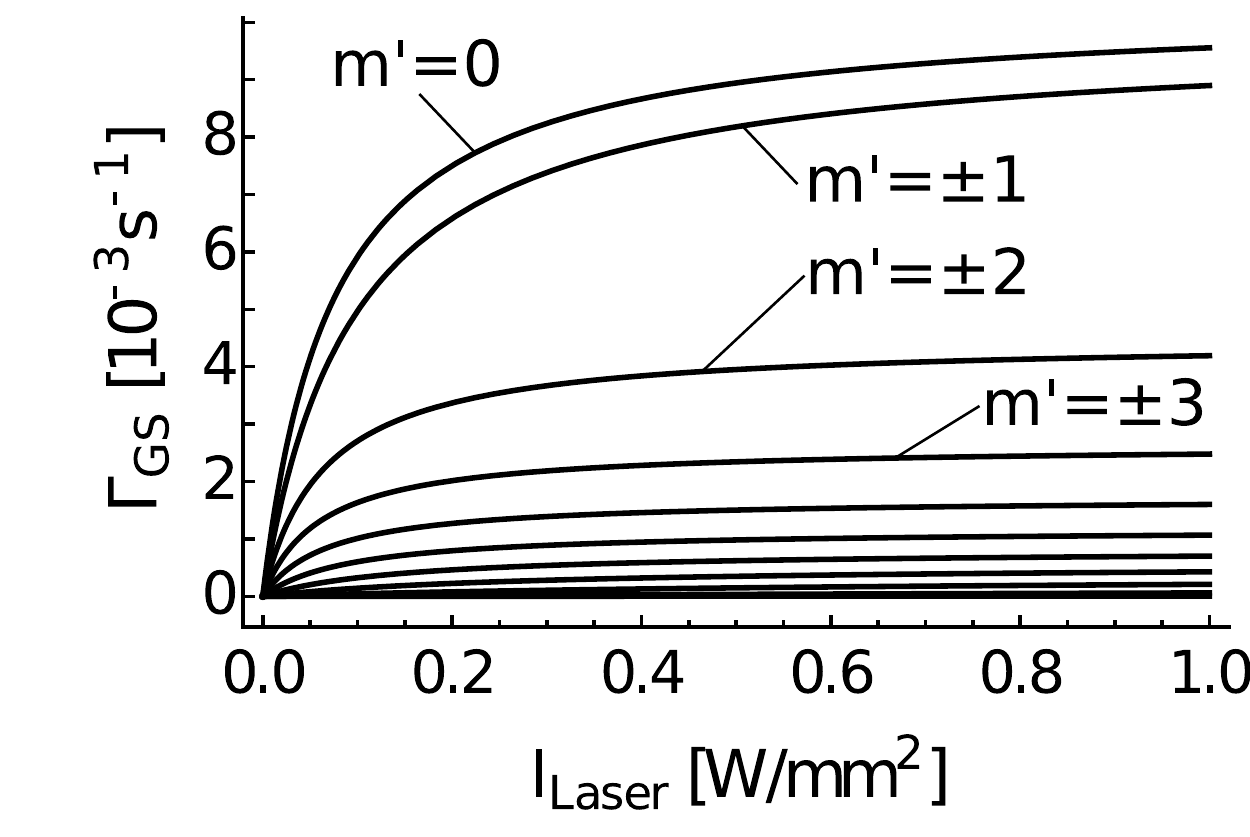}}

\caption{Ground state population rate per antiproton as a function of the laser intensity ($\pi$ polarized)  to drive the transition from the continuum to $n'=3$ (left) or to $n'=11$ (right). From those levels the ground-state is reached by spontaneous emission. The top line shows the results summed over the $m'$ states while the bottom line indicates the rate for all different $|m'|$. The positron plasma is assumed to have a density of $\unit[10^8]{cm^{-3}}$ and a temperature of $\unit[10]{K}$.}
\label{fig:srr_10_or_3} 
\end{figure}

The steady state regime in the rate equations leads to a rate toward the ground state  (cf.~Eq.~\ref{steady_state_srr_association_rate}) of 
$$  \Gamma^{\rm a}  = N_1
 \sum_{m'=-(n'-1)}^{n'-1}  (n'-|m'| ) \frac{  \Gamma^{\rm{pi}}_{m'}   \Gamma^{\rm{d}}_{m'}}{  \Gamma^{\rm{pi}}_{m'} + \Gamma^{\rm{d}}_{m'}}.$$
We deduce that the saturation intensity is reached when the  photoionization rate equals the decay rate, but the maximum possible association rates is reached  when (for all $|m'|$) $\Gamma^{\rm{pi}}_{m'}=\Gamma^{\rm{srr}}_{m'}/(N_1(n'-|m'| )) \gg \Gamma^{\rm{d}}_{m'}$ and then
$  \Gamma^{\rm a}  = N_1
\sum_{m'=-(n'-1)}^{n'-1}   (n'-|m'| )  \Gamma^{\rm{d}}_{m'}$.
Useful approximations to estimate the parameters are given 
in the appendix by Eq.~\ref{lifetime_approx} and \ref{eq_srr_kramers}.
Because of the $1/|m'|$ scaling of $ \Gamma^{\rm{d}}_{m'}$ only small values of $m'$ will dominate. Fortunately, this is also toward these low $|m'|$ states that the srr rate (or the photoionization rate $\Gamma^{\rm{pi}}_{m'}$) is the largest (cf.~Fig.~\ref{fig:radiativerecombination} in the appendix) thus the laser power required to saturate the srr of low $m'$ states is smaller.
This is also illustrated in Fig.~\ref{fig:srr_10_or_3} which indicates the ground state association rate for radiative recombination toward $n'=11$ or $n'=3$. These levels have been chosen because powerful lasers at convenient wavelengths (respectively, CO$_2$ at \unit[11]{$\mu$m} and Ti:Sa at \unit[820]{nm}) exist. On the contrary to go down to $n'=2$ would require
a UV wavelength ($\unit[365]{nm}$). Different laser polarizations are possible leading to slightly different results because the number of continuum  states that are addressed from a given $n' $ manifold is higher for $\sigma^\pm$ than for $\pi$ polarization. However, in order to avoid mixing of $m$ levels, we choose to give results only for $\pi$ polarization.

In the $n'=11$ case the saturation power is of the order of \unit[0.3]{W/mm$^2$} which can very easily be reached, but the GS association rate is only of the order of a few percent per antiproton per second. On the contrary, in the $n'=3$ case the saturation power is of the order of \unit[10]{kW/mm$^2$} which is very hard to reach, but the GS rate is of the order of one per antiproton per second. Both are well below the 3BR rates, see Eq.~\ref{3BR_rate}, but produce directly ground-state antihydrogen.

\subsubsection{Srr followed by stimulated deexcitation}

As suggested already by \cite{wolf1993laser}, enhancing the decay $\Gamma^{\rm{d}}$ of the populated $n'=11$ level by creating a stimulated deexcitation cascade down to $n''=3$ (or $n''=2$) in order to take advantage of both the higher ground-state rate and the high available laser power seems promising to enhance the prospects of srr formation. We thus investigated in more detail this scheme in order to provide the necessary laser intensities to obtain an optimal ground state formation rate.

Here again the  fact  that the srr process favors formation of low angular momentum states  (cf.~Fig.~\ref{fig:radiativerecombination})  works in favor of this scheme since such levels can be coupled by a laser to lower $n''$ levels.  Furthermore, the low $|m'|$ angular momentum states are the most numerous (degeneracy $n' - |m'|$) and, due to the diamagnetic and collisional reshuffling, they all have transitions allowed for deexcitation.

\begin{figure}
\centering
\subfloat[]{\includegraphics[width=0.8\linewidth]{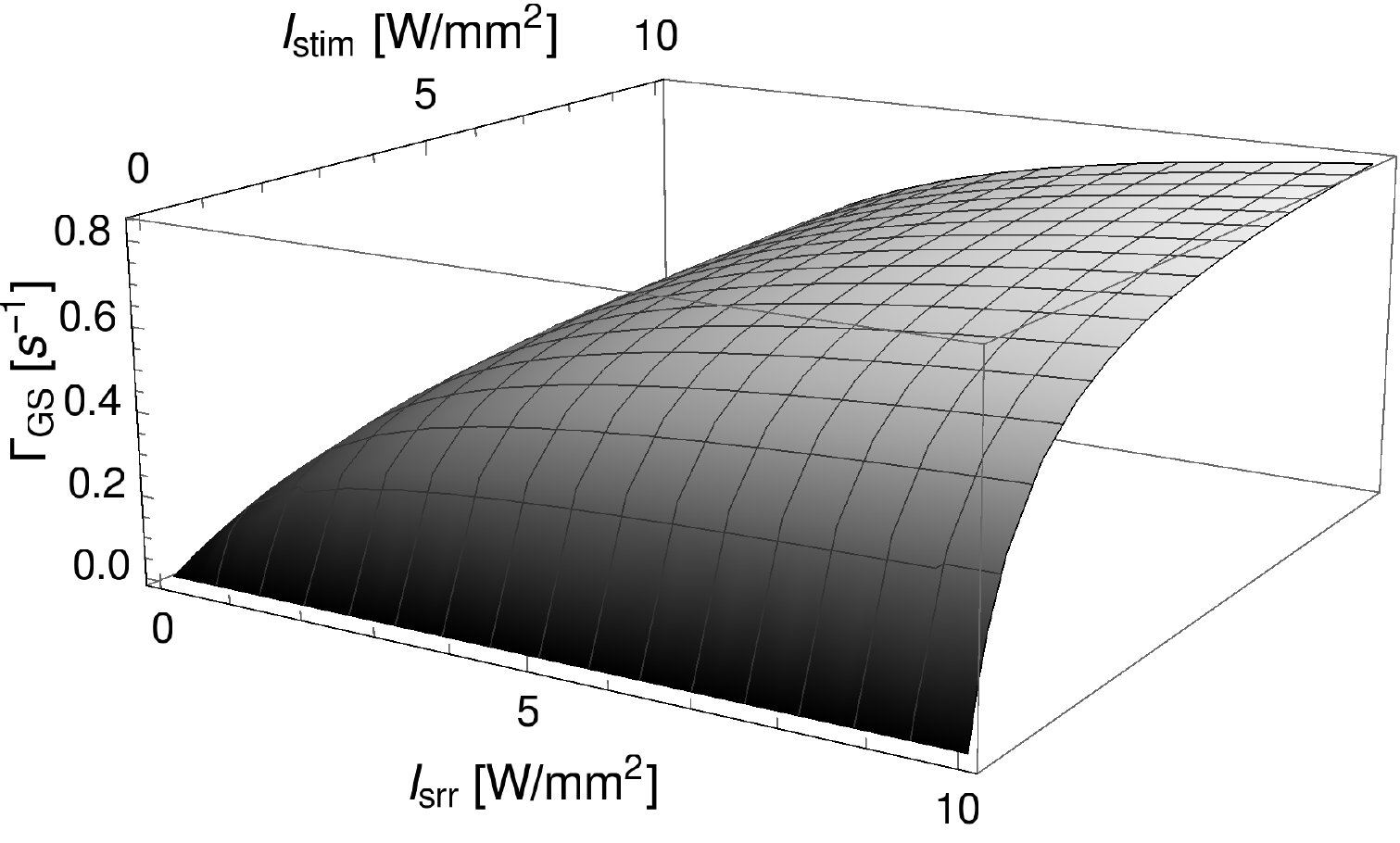}}

\subfloat[]{\includegraphics[width=0.8\linewidth]{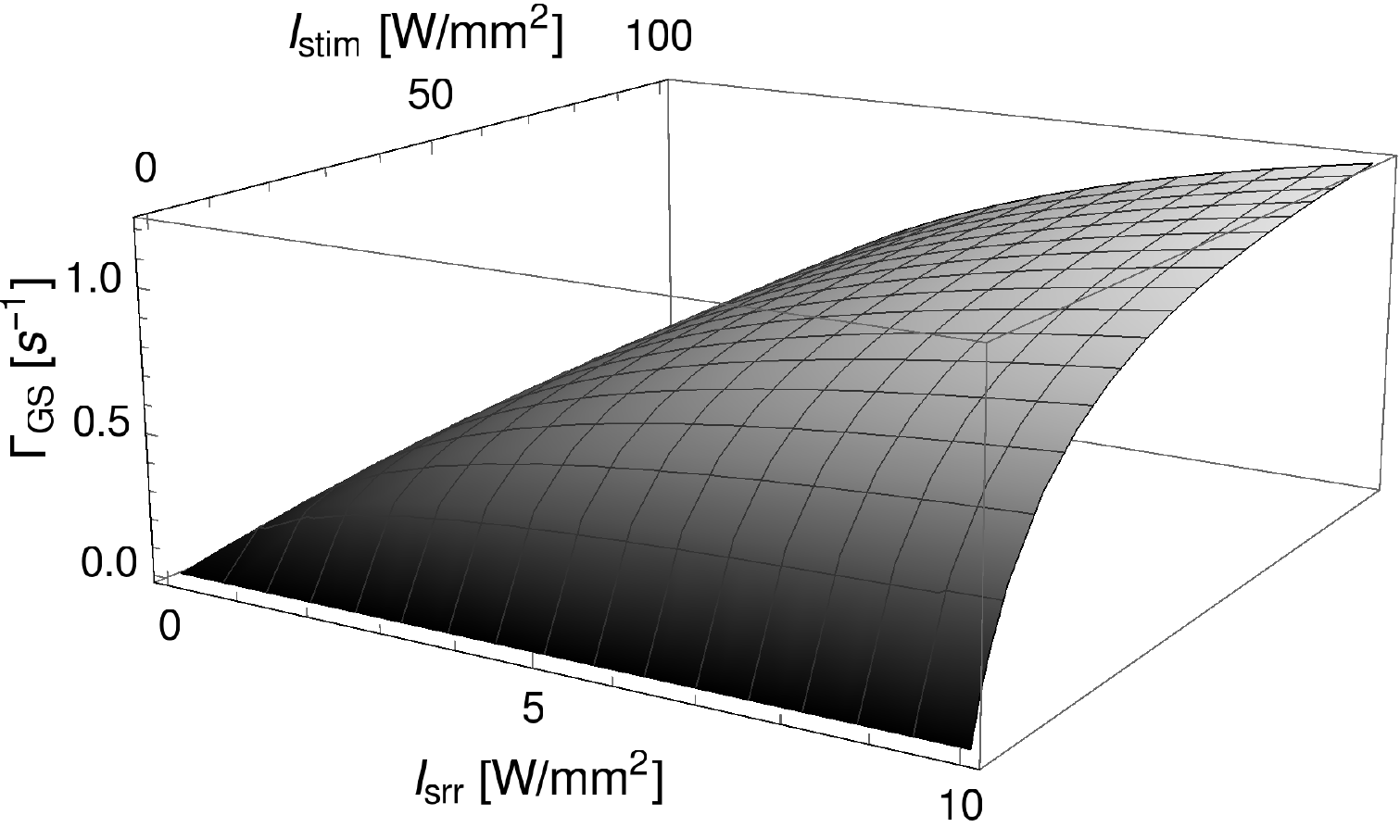}}
\caption{Ground state population rate per antiproton as a function of the laser intensity ($\pi$ polarized) to drive the transition from the continuum to $n'=11$ (srr step) and then (stimulated step) to $n''=3$ (top) or $n''=2$ (bottom). The laser is $\pi$ polarized with a FWHM bandwidth of \unit[500]{MHz}. The positron plasma has a density of $\unit[10^8]{cm^{-3}}$ and a temperature of $\unit[10]{K}$.}
\label{fig:srr_10_3} 
\end{figure}

In Fig.~\ref{fig:srr_10_3} we plot the association rate to populate the ground state  as a function of the laser powers for the srr step (continuum down to $n'=11$) as well as for the stimulated step $n'=11\rightarrow n''=3$ or  $n'=11\rightarrow n''=2$. Since this last laser has to be able to drive all transitions, we choose a FWHM laser linewidth of \unit[500]{MHz} that is sufficient to cover the diamagnetic $\sim \unit[500]{MHz}$ as well as
the  $\sim \unit[160]{MHz}$ collisional broadening. Assuming, for simplicity, a uniform laser power over all transitions (top-hat spectral profile), we can use the fully $k$-mixed formulae derived in the appendix. Going to $n''=3$, we see that we can reach a ground-state population rate of the order of $\unit[0.8]{s^{-1}}$ per antiproton with laser powers of the order of $\unit[10]{W}$ for both the srr laser ($\unit[11]{\mu m}$) and the stimulated laser ($\unit[885]{nm}$) which are within experimental reach. Going to $n''=2$ recovers the association rate given in Eq.~\ref{max_srr_rate} which is slightly above the $\unit[0.8]{s^{-1}}$ for $n''=3$, but with a laser power at $\unit[377]{nm}$ which is $\unit[100]{W}$ and thus remains difficult to achieve. Nevertheless, the results confirm the initial suggestion by Wolf \cite{wolf1993laser} (which was however made in a field-free environment while we consider the effects of the collisions and diamagnetic mixing) that a two-step srr process can lead to a significant formation rate of antihydrogen in the ground state. Using two lasers with \unit[10]{W} power, a rate of the order of $\sim \unit[1]{s^{-1}}$ per antiproton is obtained which means that with the typical $\unit[10^6]{}$ antiprotons trapped and within a plasma interaction time of a millisecond, roughly 1000 atoms in ground-state can be produced at a time.

\subsubsection{Discussions}

The high laser intensity required, especially for the single-step srr, can raise the question of using pulsed stimulated radiative recombination that would have also the advantage to produce antihydrogen in a pulsed manner. This can be useful for further manipulations using pulsed fields and for experiments requiring a time-of-flight measurement \cite{doser2018aegis}.
Due to the lower duty cycle, the production rate would be smaller, however the saturated power will be very easily reached and direct stimulation down to the $2p$-state would be feasible. 
For a short pulse we can neglect the spontaneous emission terms in the rate equation; the population of a srr level i thus evolves as $N_1 (1-e^{- \Gamma^{\rm pi}_{\rm{i}} t})$ leading to saturation when the photoionization equals the laser pulse time.
An interesting case can be when the spontaneous emission or stimulated deexcitation depopulates the formed levels of the $n'$-manifold  before the arrival of the next laser pulse such that photoionization by the next pulse is avoided. 
As shown in Eq.~(\ref{eq_srr_kramers}), the srr cross-sections are negligible  for $|m'| \gtrsim 3 n'^{0.7}/2$. Thus, a saturation pulsed srr laser can  populate $\sim 3 n'^{1.7}$ levels of the $n'$ manifold (among the $n'^2$ available).
For instance, every pulsed stimulated decay down to $n'=36$  (having a  convenient wavelength for a CH$_3$OH laser at \unit[118.8]{mm}) will populate almost 1300 levels each of them with population $N_1$. Those levels could then be deexcited using the tools developed before with almost an order of magnitude faster dynamic since the initial state is now specifically targeted and thus the number of states populated is smaller than in the usual formation mechanisms.

We finally would like to note that several effects might improve the estimations and results presented above. 
For example, we have previously neglected the quantization of the cyclotron frequency in the continuum, but in fact (quasi-) Landau resonance, separated by $3/2$ the cyclotron frequency, or similar types of ro-vibrating structure are present and could be used to enhance the process by a large factor  \cite{friedrich1989hydrogen,holle1986diamagnetism,1991PhRvL..66..141D,holle1988quasi}.
Additionally, the three body recombination, in the often quoted experimental conditions, would compete with the srr process and populate several Rydberg $n'$ levels \cite{radics2014scaling}.
The positron continuum can therefore be seen as being populated down to an energy of $\sim (1~\rm{to}~4) k_B T$. Therefore, by detuning all lasers to this bottleneck, the stimulated radiative recombination can be enhanced by a Boltzmann factor of $e^1\sim 3$ (\cite{wolf1993laser} even uses population down to an energy of $4 k_B T$ so an enhancement of $e^4\sim 45$). We however note that the pure three-body-recombination scales in positron density and temperature as $n_{e^+}^2 T^{-4.5}$ whereas the srr rate scales as $n_{\rm{e^+}}Q^{-1}=n_{\rm{e^+}}T^{-1.5}$ so that different experimental conditions would lead to different possible enhancements.

\section{\label{s:concl}Conclusions}
This work addressed the long-standing issue of the deexcitation of antihydrogen atoms formed in Rydberg states to allow measurements of their properties in ground-state. In particular, experiments aiming at forming a beam of antihydrogen atoms require a prompt deexcitation, the lack of which so-far hindered the production of a useful beam. Trap experiments could also benefit from a rapid deexcitation mechanism which could lower the final temperature of the trapped antihydrogen and enhance the trapping efficiency \cite{coolingpaper}. The presence of high angular momentum states prevents the use of a single laser for this purpose thus we propose to couple, via THz and/or microwave light, a large number of states in the high Rydberg region (around $n\sim30$) which can then be depopulated via a single laser down to low lying states (around $n'=3$).
A key point is  that the characteristic deexcitation time is fundamentally limited by the number of states addressed.  This observation led to the correction  \cite{PhysRevA.101.019904} of a previously published \cite{COM181} result which was neglecting the repopulation of the initial manifold. A final dissipative spontaneous process is required to drive the population down and should be as fast as possible, hence the choice of low-lying end states, the $2p$ level being the optimal choice to maximize the overall ground-state population rate (but not necessarily the optimal choice in terms of experimental feasibility).

Practically, based on the best available information on the antihydrogen level distribution for a given experimental condition, a choice on the manifolds and the number of states addressed would have to be made which in turn would determine the light necessary to couple and deexcite the states and the minimum achievable deexcitation time. 
The technique described has the advantage to be versatile and therefore applicable to different formation mechanisms (3BR or CE). The coupling of many states also allows to address a larger initial state distribution as present after the 3BR formation process however limited in the high-$n$ region to $n\sim35$ due to losses via excitation and ionization channels.
In this article we made the choice of addressing a distribution between the $n= 20$ and $n=35$ manifolds which led to a close to unity deexcitation in $\sim \unit[50]{\mu s}$ which is rapid enough to target the atoms at their formation point if their temperatures are in the few tens of Kelvin range.
We note that such $\mu s$ timescale deexcitation mechanism could provide a time information necessary for time-of-flight- based velocimetry measurements for diagnostics in beam experiments relying on ``continuous'' formation processes.

We showed as well that a stimulated deexcitation coupled to a stimulated recombination formation process can significantly enhance the number of antihydrogen atoms produced in ground-state with reasonable laser powers.
The main difference between the srr process and the bound-bound stimulated deexcitation process, is that the initial population of a single state does not anymore inversely scale  with the number of states coupled to each other via THz or microwave, but is given by the  plasma  degeneracy  parameter (phase  space  density) 
$1.3\times10^{-9} = \frac{n_e^+}{\unit[10^8]{cm^{-3}}} \left( \frac{T}{\unit[10]{K}} \right)^{-1.5} $ and that this initial population is always refilled by collisions. We showed that in the usual experimental conditions used for three-body-recombination the laser-stimulated (followed by stimulated deexcitation) formation rate per antiproton would still be about two orders of magnitude lower than for the three-body-recombination. However, in the former case all atoms are in ground-state compared to the tiny fraction in the case of pure three-body-recombination. We also noted that due to the different scalings of the rates with respect to the positron temperature and density, the stimulated recombination is less affected by an increased positron temperature and density, thus being in this respect easier to achieve and more robust to changes in experimental conditions.
We finally note that the antiprotons in the plasma, and thus the formed antihydrogen atoms, have finite velocities. The Doppler broadening originating from their motion was not taken into account in the manuscript and would lead in general to the use of spectrally broader lasers. Alternatively, deexcitation lasers with sharp linewidths can be used as a velocity selector which can be of interest for the formation of an antihydrogen beam.

\section*{Acknowledgements}
This work has been sponsored by the Wolfgang Gentner Program of the German Federal Ministry of Education and Research (grant no. 05E15CHA, university supervision by Norbert Pietralla). It was supported by the German Academic Scholarship Foundation and the Fond Unique Interminist\'eriel (IAPP-FUI-22) COLDFIB.

\appendix

\section{Stimulated radiative recombination}\label{s:appendixssrr}
This appendix goes into the details of the srr process by illustrating the mechanism through three different approaches (standard srr theory, analogy with photoassociation and rate equations) which we find useful to grasp the physics at play and verify the consistency of the different approaches.

\subsection{Milne relations with photoionization}

In a srr process, a laser associates (label (a)) a positron from the continuum level with an antiproton to form a bound level that can be photoionized (label (pi)), decay or be detected (label (d)).
The  stimulated radiative recombination (srr) rate per antiproton from the single and non degenerate
$\rm{j}$  antiproton state
 into a bound state $\rm{i}$  is given by \cite{bauche2015atomic}:
 
\begin{equation}
\Gamma^{\rm{srr}}_{\rm{j} \rightarrow \rm{i}} = n_{e^+} \int_0^\infty \frac{I(\nu)}{ h \nu}  \sigma^{\rm srr}_{\rm{j} \rightarrow \rm{i}} v f(v) {\rm d} v,
\label{srr_eq_rate}
\end{equation} 
where the velocity distribution $f(v)$ will, in our case, be 
a Maxwellian distribution  $f(v) = 4 \pi v^2  \left( \frac{m}{2 \pi k_B T} \right)^{3/2} e^{-\frac{1}{2} m v^2 /k_B T}$.
$\sigma^{\rm srr}_{\rm{j} \rightarrow \rm{i}}$ is
the stimulated radiative recombination cross-section (because stimulated recombination is a three-body process, it has
the dimension of the square of a surface) and
$I = \int I(\nu) {\rm d} \nu $  is the  laser irradiance. 
For a laser of frequency $\nu$ there is a relation between $\nu$ and $v$ through the positron energy above the ionization threshold:
  $E_0=  \kappa^2 R_y = h  \nu - R_y/n'^2 = \frac{1}{2} m v^2$.
The laser-induced intra-continuum 
(free-free) transitions 
 can be neglected
 \cite{neumann1983laser} and the
srr process
$\Gamma^{\rm{srr}}_{\rm{i}}  =  \sum_{\rm{j}} \Gamma^{\rm{srr}}_{\rm{j} \rightarrow \rm{i}} $  toward level $\rm{i}$ only 
competes with the spontaneous emission and the reverse
photoionization rate
 from level $\rm{i}$ being $\Gamma^{\rm{pi}}_{\rm{i}} =\sum_{\rm{j}} \Gamma^{\rm{pi}}_{\rm{i} \rightarrow \rm{j}}$ where
\begin{equation*}
  \Gamma^{\rm{pi}}_{\rm{i} \rightarrow \rm{j}}=
 \int  \frac{I( \nu)}{ h   \nu}  \sigma^{\rm pi}_{\rm{i} \rightarrow \rm{j}}(\nu)   {\rm d}  \nu.
\end{equation*}

The photoionization cross-sections $\sigma^{\rm pi}$, the  $\sigma^{\rm rr}$ one for radiative recombination due to spontaneous emission and 
$\sigma^{\rm srr}$ for the  stimulated  radiative recombination
are linked through the detailed balance and microreversibility relation
\begin{equation*}
 \frac{c^2 }{8 \pi \nu^2} \sigma^{\rm rr}_{\rm{j} \rightarrow \rm{i}}(v,\nu) =  \sigma^{\rm srr}_{\rm{j} \rightarrow \rm{i}}(v,\nu) = \frac{h^2}{8 \pi m^2 v^2 } \frac{g_{\rm i}}{g_{\rm j}} \sigma^{\rm pi}_{\rm{i} \rightarrow \rm{j}}(\nu)
\end{equation*}
where, to be more general, we have added $g_{\rm{i}}$ and $g_{\rm{j}}$  possible degeneracy numbers (for instance ($n',l'$) is degenerated $2(2l'+1)$ times because of the electron spin). 
These so called
Milne's relations can be obtained
by equating photoionization and stimulated plus spontaneous emission rates in the Saha-Boltzmann thermal equilibrium ($n_{e^+} \Lambda_T^{3} n'^2 e^{R_y/n'^2 k_B T}$) and under Planck irradiance (where spontaneous emission is a factor $\bar n = \frac{1}{e^{ h  \nu/k_B T}-1}$ smaller than the stimulated one).
For non degenerate levels (thus when the electron spin is included) such as ${\rm{i}} = |n' k' m' m'_s \rangle$ and ${\rm{j}} = |E k m m_s \rangle$ ($g_{\rm{i}}=g_{\rm{j}}=1$), all the previous formulae lead to the fundamental relation
$$\Gamma^{\rm{srr}}_{\rm{j} \rightarrow \rm{i}} = n_{e^+} Q_T^{-1}   e^{-E_0/k_B T} \frac{1}{2} \Gamma^{\rm{pi}}_{\rm{i} \rightarrow \rm{j}} $$
with $Q_T=\Lambda_T^{-3}$, where 
 $\Lambda_T= \frac{h}{\sqrt{2\pi m k_B T}}$ is the thermal de Broglie wavelength.

\subsection{Rate equations}

The association rate is given by the rate to populate the ground state
through the decay, at a rate $\Gamma^{\rm{d}}_{\rm{i}}  $, of a level ${\rm{i}} = | n' k' m' m'_s\rangle$ that is populated by srr from the continuum states j, but is also  photoionized.  This leads to the rate equations illustrated in Fig.~\ref{fig:srr_rates} (left). 

An important assumption of the srr models is that the collisions in the continuum are faster than the srr transfer and that the amount of transfer is negligible. Therefore, the continuum is seen as being in a steady state and the population of individual levels is constant
$N_{\rm{j}} =  N_c$. 
The steady state of the rate equations shows that $N_{\rm{i}}$, the population of level i, is constant and $\Gamma^{\rm a}_{\rm{i}} =  \Gamma^{\rm d}_{\rm{i}} N_{\rm{i}}  = \frac{   
\Gamma^{\rm{srr}}_{\rm{i}}   \Gamma^{\rm{d}}_{\rm{i}}}{  \Gamma^{\rm{pi}}_{\rm{i}} + \Gamma^{\rm{d}}_{\rm{i}}}$.
This leads to 
$N_c= n_{e^+} Q_T^{-1}   e^{-E_0/k_B T}/2$ and $\Gamma^{\rm{srr}}_{\rm{j} \rightarrow \rm{i}} = N_c \Gamma^{\rm{pi}}_{\rm{i}\rightarrow \rm{j}} $.

In summary, the association  rate 
$\Gamma^{\rm a} =   \sum_{\rm{i}=1}^{g_{\rm i}}
\Gamma^{\rm a}_{\rm i}$, or more precisely, the  decay measured population into bound states, degenerated $g_{\rm{i}}$ times, is  given by
\begin{equation}
    \Gamma^{\rm a}  =
 n_{e^+} Q_T^{-1}   e^{-E_0/k_B T} \frac{1}{2}
 \sum_{\rm{i}=1}^{g_{\rm{i}}} \frac{  \Gamma^{\rm{pi}}_{\rm{i}}   \Gamma^{\rm{d}}_{\rm{i}}}{  \Gamma^{\rm{pi}}_{\rm{i}} + \Gamma^{\rm{d}}_{\rm{i}}}.
 \label{eq_srr_final}
\end{equation}

If we now sum over the electron spin and use the fact that
only $m_s = m'_s$ are authorized transitions ($\pi$ transitions),  the formula becomes $\Gamma^{\rm{srr}}_{E l m  \rightarrow n' l' m' } = 2 \Gamma^{\rm{srr}}_{E l m m_s \rightarrow n' l' m' m_s}$ and so
$  \Gamma^{\rm a}  = N_1
\sum_{\rm{i}} \frac{  \Gamma^{\rm{pi}}_{\rm{i}}   \Gamma^{\rm{d}}_{\rm{i}}}{  \Gamma^{\rm{pi}}_{\rm{i}} + \Gamma^{\rm{d}}_{\rm{i}}}$, but now with states $\rm{i}$, such as $|n k m \rangle$, that do not have anymore the electron spins. The population of each of these continuum states  is $N_1 = 2 N_c = n_{e^+} Q_T^{-1}   e^{-E_0/k_B T}$.

\subsection{Link with photoassociation}

An interesting analogy can be done with the microwave or photoassociation process in which two colliding atoms absorb a photon to form a molecule.
The advantage of this approach is that a detailed theory has been developed \cite{weiner1999experiments,jones2006ultracold,thompson2005ultracold}. The photoassociation rate (more precisely the detected rate of the population of the bound  level) per antiproton is given, for a narrowband laser of frequency $\nu$, by (using field free notations)
$$
\Gamma^a_{\rightarrow n' l' m' m'_s} =n_{e^+} \times\frac{1}{h Q_T} \int  e^{-E/k_B T} \frac{1}{2} \sum_{m_s} \sum_{l m }   |S|^2 \rm{d}E.
$$
Here, $n_{e^+}$ is the positron density, $Q_T = \left( \frac{2\pi \mu k_B T}{h^2} \right)^{3/2}$ the translational partition function ($\mu \sim m$ is the reduced mass between positron and antiproton) and, for a collision energy $E$, the $S$-matrix element is
$|S|^2 = \frac{ \hbar \Gamma^{\rm{b}} \hbar \Gamma^{\rm{d}} }{(E-E_0)^2 + (\hbar (\Gamma^{\rm{b}}+\Gamma^{\rm{d}})/2)^2 }$  where we have neglected the light shift.   
$\Gamma^{\rm{d}}$ is the decay (or detection) rate of the bound level (due for example to collision, spontaneous emission, ejection out of the laser zone, etc.).  The stimulated rate $\Gamma^{\rm{b}} $ toward the bound level $|n' l' m' m'_s\rangle$   is given by Fermi's Golden rule $\hbar \Gamma^{\rm{b}} = 2\pi |\langle n' l' m' m'_s | e {\bm r} . \bm{E_{\rm{L}}}/2 |E l m m_s\rangle |^2$ for a laser electric field $E_{\rm{L}}^2=2I/c \varepsilon_0$. 
The absorption rate equals the stimulated emission one
$ \Gamma^{\rm{pi}}_{n' l' m' m'_s \rightarrow E l m m_s} = \Gamma^{\rm{b}} $.
If we suppose that the positron continuum is larger than any Lorentzian atomic linewidth, we find that only the collision energy $E_0=  \kappa^2 R_y = h  \nu - R_y/n'^2 $ matters and that the photoassociation rate at resonance is 
\begin{equation}
n_{e^+} \frac{h^2}{(2\pi \mu k_B T)^{3/2}}   e^{-E_0/k_B T} \frac{1}{2} \sum_{m_s} \times\sum_{l m }   \frac{ 2\pi \hbar \Gamma^{\rm{b}} \hbar \Gamma^{\rm{d}} }{\hbar (\Gamma^{\rm{b}} + \Gamma^{\rm{d}})} \label{PA_rate}.
\end{equation}
This is exactly the rate
calculated before (cf.~Eq.~\ref{eq_srr_final}) which points out the consistency of both methods.
The advantage of the photoassociation picture is that it directly incorporates the saturation in the $S$ matrix expression.

\subsection{Effective fully mixed system}

\begin{figure}
\centering
\includegraphics[width=0.8\linewidth]{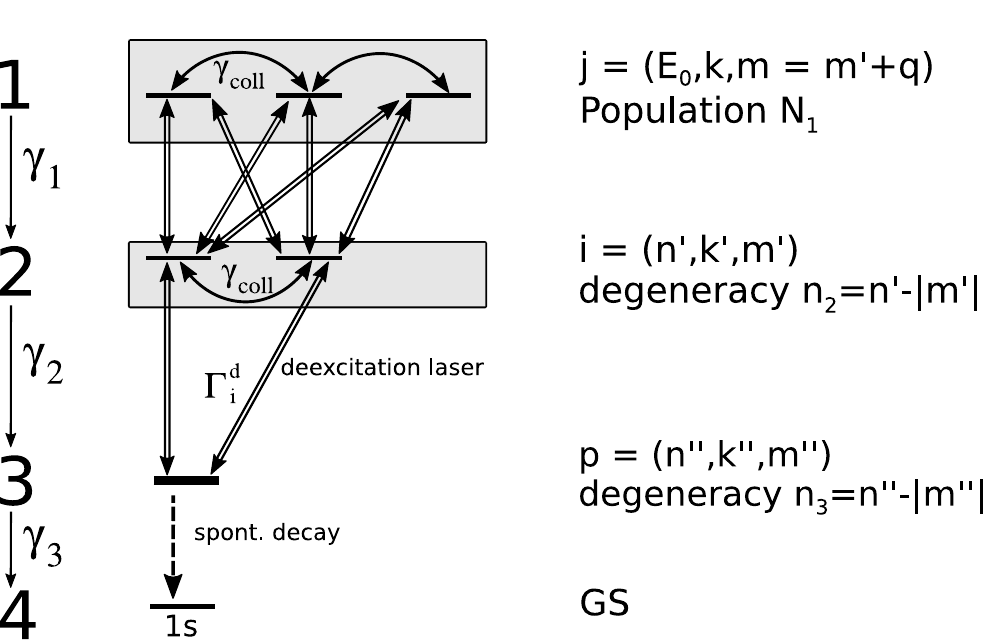}
\caption{Level system due to perfect $k$ mixing created by collisions that equidistribute the population in the ($n',m'$) manifold. The simplified system with 4 levels is indicated on the left of the figure.}
\label{fig:srr_4levels_equations} 
\end{figure}

We have seen the consistency between the photoassociation,  srr and simple rate equation models. We now extend the simple rate equation model in order to include the collisional mixing and a stimulated emission step from the srr targeted levels.
As discussed in section \ref{s:collis}, due to the diamagnetic shift in the presence of a $\bm B$ field and collisions within the plasma, the $k$ levels will be fully mixed whereas $n$ and $m$ will stay rather well defined.
We can thus study in an isolated way the srr toward an ($n',m'$) manifold. It will be coupled to $m=m'+q$ levels in the continuum depending on the polarization $q$ of the srr laser. We can also, using a stimulated laser with polarization $q'$, deexcite these levels toward an ($n'',m''$) manifold that spontaneously decays toward the ground state. We illustrated the rate equations in Fig.~\ref{fig:srr_rates} (right) using the notation given in Fig.~\ref{fig:srr_4levels_equations}.

The srr rate is given by $\Gamma^{\rm srr}_{\rm j\rightarrow \rm i} = \sum_{\rm j}
\gamma_{1;\rm i,j} N_{1,\rm j}$ which leads to the following rate equations:
\begin{widetext}
\begin{eqnarray*}
  N_{1,\rm j} &=& n_{e^+} Q_T^{-1}   e^{-E_0/k_B T} \\
  \frac{\rm{d} N_{2,\rm i}}{\rm{d} t } & = &   \sum_{\rm j} \gamma_{1;\rm i,j} (N_{1,\rm j} - N_{2,\rm i} ) -    \sum_{\rm p} \gamma_{2;\rm i,p} (N_{2,\rm i} - N_{3,\rm p} ) + \gamma_{\rm coll} \sum_{\rm i'} (N_{2,\rm i'}-N_{2,\rm i}) \\
    \frac{\rm{d} N_{3,\rm p}}{\rm{d} t } & = &   \sum_{\rm i} \gamma_{2;\rm i,p} (N_{2,\rm i} - N_{3,\rm p} ) -   \gamma_{3;\rm p} N_{3,\rm p}   + \gamma_{\rm coll} \sum_{\rm p'} (N_{3,\rm p'}-N_{3,\rm p}) \\
    \frac{\rm{d} N_4}{\rm{d} t } & = &  \sum_{\rm p}  \gamma_{3;\rm p } N_{3,\rm p}
\end{eqnarray*}
\end{widetext}
The decay rate $\gamma_{3;\rm p }$ from the ($n'',m''$) manifold is not the decay rate directly toward the ground state, but since eventually the entire population will reach the ground state after another cascade this will also be the rate of population of the ground state ($N_4$).

The collisional rate $\gamma_{\rm coll}$ (typically one per nanosecond for the $k$-mixing) being much faster than any other rate, we chose to have the same notation for the collisional rate within the i and j states. Indeed, the net result is that a quasi stationary state is reached with an equidistribution between the states ($N_{2,\rm i'}=N_{2,\rm i}$ and $N_{3,\rm p'}=N_{3,\rm p}$). 
We will note the population of individual levels (summed over the electron spin) as:  $N_1= n_{e^+} Q_T^{-1}   e^{-E_0/k_B T}$  for the continuum,
$N_2$ (that equals all $N_{2,\rm i}$) for the ($n',m'$) manifold (with degeneracy $n_2=n'-|m'|$) and  $N_3$ (that equals all $N_{3,\rm p}$) for the ($n'',m''$)  manifold (with degeneracy $n_3=n''-|m''|$). 
By summing over $\rm i$ and $\rm p$ we obtain, for an evolution which is slower than the collisional rates, the following rate equations:
  \begin{eqnarray*}
  N_1&=& n_{e^+} Q_T^{-1}   e^{-E_0/k_B T} \\
  \frac{\rm{d} N_{2}}{\rm{d} t } & = & - (n_3 \gamma_2 + n_1 \gamma_1) N_{2} + \gamma_1 n_1 N_{1} + n_3 \gamma_2  N_3 \\
    \frac{\rm{d} N_3}{\rm{d} t } & = & n_2 \gamma_2 N_{2}  - n_2 \gamma_2 N_3 -\gamma_3 N_3 \\
    \frac{\rm{d} N_4}{\rm{d} t } & = &   n_3 \gamma_3 N_3 \
\end{eqnarray*}
with the average rates: $  \gamma_1 =  \frac{1}{n_1 n_2} \sum_{\rm i,j} \gamma_{1;\rm i,j}$,
$ \gamma_2 =  \frac{1}{n_2 n_3} \sum_{\rm i,p} \gamma_{2;\rm i,p} $ and 
$ \gamma_3 =  \frac{1}{ n_3} \sum_{\rm p} \gamma_{3;\rm p} $.

This leads to a steady state rate of $N_3  = N_1 \frac{1}{1+ \gamma_3\left( \frac{1}{n_2 \gamma_2} + \frac{n_3 }{n_1 n_2  \gamma_1 } \right)}$. Consequently,
\begin{equation}
      \Gamma^{\rm a}  =     n_3  \gamma_3 N_3 = n_3  \gamma_3  N_1 \frac{1}{1+ \frac{n_3 \gamma_3}{n_2}\left( \frac{1}{ n_3 \gamma_2} + \frac{1 }{n_1  \gamma_1 } \right)}.
      \label{steady_state_association_rate}
\end{equation}
This simple formula restores all results found in this article (also in the case of stimulated deexcitation of bound levels since the srr is mainly a stimulated decay from levels that have $N_1$ population whereas for bound levels $N_1=1/N$, $N$ being the number of states coupled to each other. The additional difference is that the initial levels of the srr are always repopulated by collisions). The main findings are:
\begin{itemize}
    \item The last levels toward which the decay is stimulated should be the ones that spontaneously decay the fastest ($n_3 \gamma_3$ maximum), so ideally the $n=2$ level.
    \item The final rate will be limited by the slowest rate in the cascade: so either the srr step with rate $n_1 \gamma_1$ or the stimulated deexcitation one with rate $n_3 \gamma_2$.
    \item The most efficient case is when all rates are equal $ n_3 \gamma_2 = n_1  \gamma_1 $.
    \item The maximum possible association rate is simply given by the full transfer between non-degenerate level population $N_1$ toward the last level (so $N_3=N_1$) that decays so
   $  \Gamma^{\rm a}_{\rm max}  =  n_3  \gamma_3  N_1$.
\end{itemize}

For a pure srr process without any extra stimulated laser, so with $N_2$ playing the role of the continuum $N_1$, we would have found as a steady state 
   \begin{equation}
      \Gamma^{\rm a}  =     n_3  \gamma_3 N_3 = n_3  \gamma_3  N_1 \frac{1}{1+ \frac{ \gamma_3}{n_2 \gamma_2}}.
      \label{steady_state_srr_association_rate}
\end{equation}
That is $  \Gamma^{\rm a}  =   N_1  n_3  \gamma_3 \frac{n_2 \gamma_2}{n_2 \gamma_2+ \gamma_3 }$ which is exactly Eq.~\ref{eq_srr_final} with $\Gamma^{\rm{d}}_{\rm i} = \gamma_3$ and
$ \Gamma^{\rm{pi}}_{\rm i} = n_2 \gamma_2$.

\subsection{Rates for full $k$-mixing}

We can now specify the rate equations and notations for our specific case:
 $\gamma_2 = \frac{1}{n_3 n_2} \sum_{k' \geq |m'|}  \sum_{k''  \geq |m''|} \gamma_{n' k' m' \rightarrow n'' k'' m'-q'}$, where 
$\gamma_{n' k' m' \rightarrow n'' k'' m''} =
 \frac{2 I e^2 }{\hbar^2 \epsilon_0 c \Gamma_{\rm L}}
\left| \langle n' k' m' | r^{(q)}| n'' k'' m'''  \rangle \right|^2$ is the stimulated rate toward the $m''=m'-q$ state due to a resonant laser of intensity $I$ and FWHM linewidth of $\Gamma_{\rm L}$. Because of the unitary transformation to go from the $k$ basis to the $l$ one, induced by the diamagnetic term, the sum can also be written as
 $$\gamma_2 
=  
\frac{1}{n_3 n_2} \sum_{l' \geq |m'|}  \sum_{l''  \geq |m''|} \gamma_{n' l' m' \rightarrow n'' l'' m'-q'}$$
with
$\gamma_{n' l' m' \rightarrow n'' l'' m''} =
 \frac{2 I e^2 }{\hbar^2 \epsilon_0 c \Gamma_{\rm L}}
\left| \langle n' l' m' | r^{(q)}| n'' l'' m'''  \rangle \right|^2$ that we calculate using standard radial overlap formulae \cite{COM181}.

Similarly, the spontaneous emission rate from a (collisionally reshuffled level of the ($n'',m''$) manifold) is given by $$\gamma_3 =\Gamma^{\rm spon}_{n'' m''} = \frac{1}{n_3} \sum_{l'' \geq |m''|}  \sum_{n'''} \sum_{l''' \geq |m''|} \Gamma^{\rm spon}_{n'' l'' m'' \rightarrow n''' l''' m'''}.$$

We find a very good approximation 
 $\Gamma^{\rm{d}}_{m''}=\Gamma^{\rm spon}_{n'' m''}$ corresponding to a lifetime of 
\begin{equation}
    \frac{1}{\Gamma^{\rm{d}}_{m''}}\sim \unit[1]{\mu s} \times (n''/10)^4 |m''| 
    \label{lifetime_approx}.
\end{equation}  
The approximations are valid within 10\%
for $|m''|>0$, whereas, for $m''=0$, the fit becomes $\unit[1]{\mu s} \times (n''/10+0.04)^4$. 

\subsection{Srr and photoionization rates} 
\label{sec: Srr and photoionization rates}

\begin{figure}
\centering
\subfloat[\label{fig:9a}]{\includegraphics[width=0.5\linewidth]{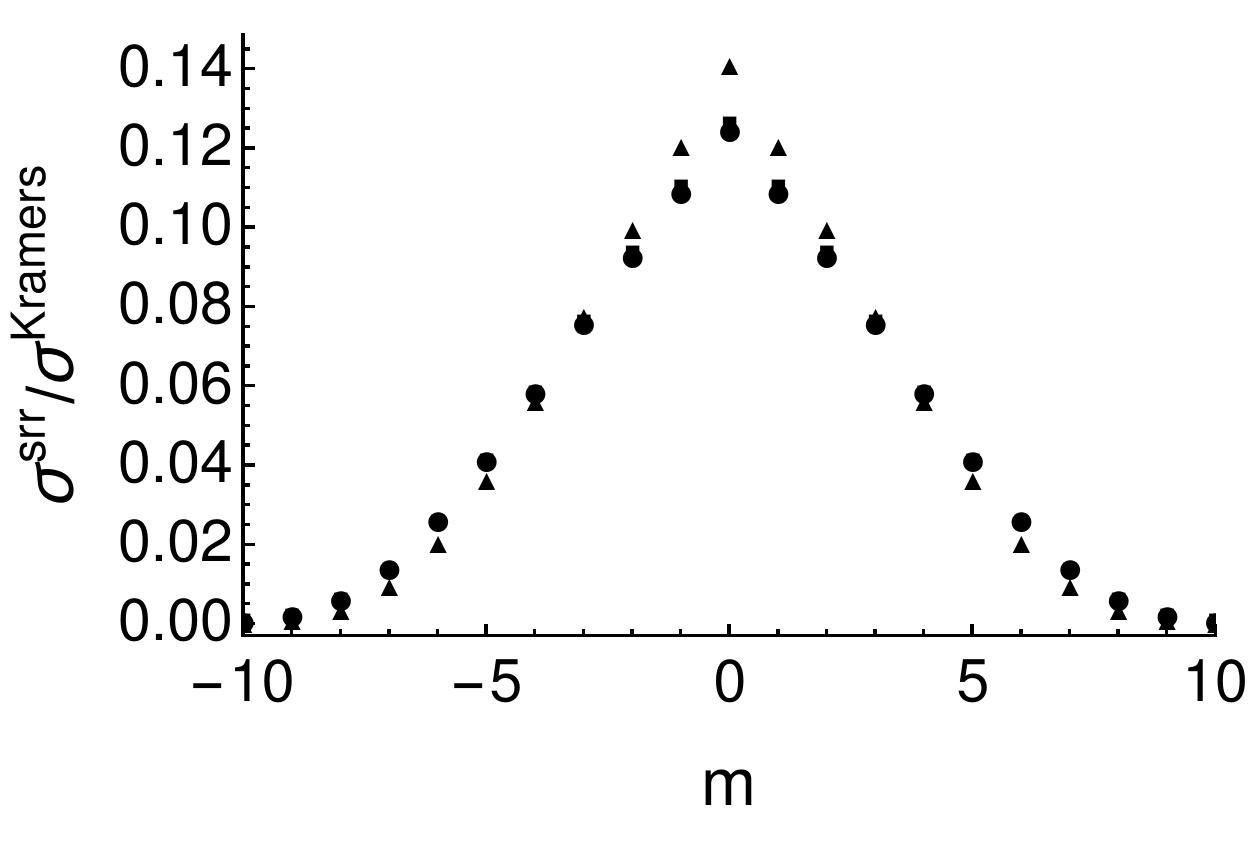}}
\subfloat[\label{fig:9b}]{\includegraphics[width=0.5\linewidth]{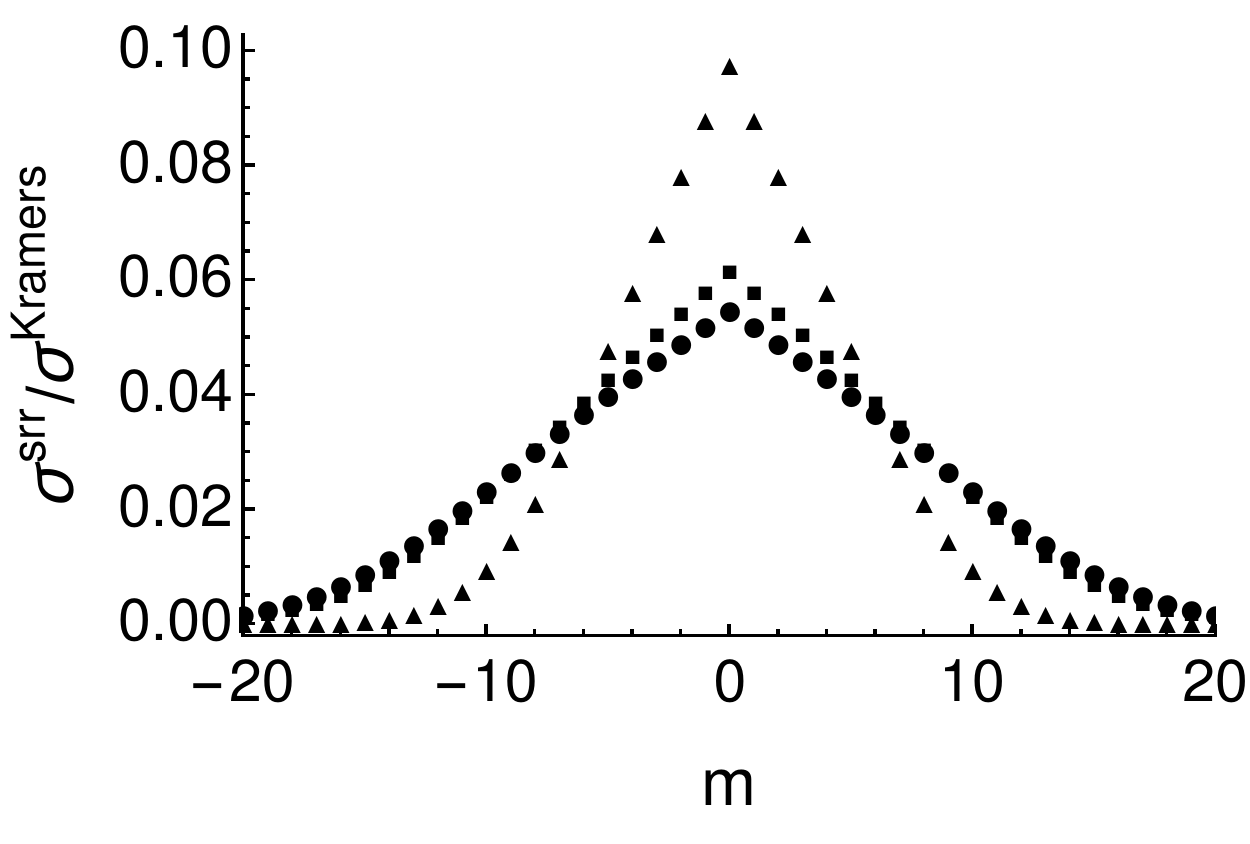}}
\caption{Stimulated radiative recombination cross-section for unpolarized light toward $n'=11$ (left) and $n'=36$ (right) at an energy of $k_B T$ in units of the (classical) Kramers' approximation $\sigma^{\rm srr, Kramers}_{E= \kappa^2 R_y \rightarrow n'} = 
 \frac{4 h^4 \alpha^{-3} }{ 3 \sqrt{3} \pi^2 m_e^4  c^4}
\frac{n'^3}{\kappa^2(1+n'^2 \kappa^2)^3} \approx
\unit[6.96 \times 10^{-42} ]{m^2} \frac{n'^3}{\kappa^2(1+n'^2 \kappa^2)^3}$ \cite{bethe2012quantum} and in a fully mixed $k'$ level environment. We give results for $\unit[10]{K}$ (circular markers), $\unit[100]{K}$ (square markers) and $\unit[1000]{K}$ (triangular markers).
For a narrowband laser the link with the srr rate is given by
 $ \Gamma^{\rm{srr}}_{\rm{j} \rightarrow \rm{i}} = n_{e^+}  \frac{I}{ \nu m}  \sigma^{\rm srr}_{\rm{j} \rightarrow \rm{i}}  f(v) $ (with $ h  \nu - R_y/n'^2 = \frac{1}{2} m v^2$).}
\label{fig:radiativerecombination}
\end{figure}

Finally, the reshuffled photoionization rates are also equal and will be noted
$$
\gamma_1 = 
 \sum_{l'\geq |m'|} 
\Gamma^{\rm pi}_{ n' l' m' \rightarrow E l'+1 m'+q }+\Gamma^{\rm pi}_{n' l' m' \rightarrow E l'-1 m'+q}.
$$

For a narrowband laser (meaning with a spectral bandwidth much smaller than the continuum one of $k_B T$), we can calculate $\Gamma^{\rm{pi}}_{\rm{i} \rightarrow \rm{j}}=
\frac{I}{ h   \nu}  \sigma^{\rm pi}_{\rm{i} \rightarrow \rm{j}}$.
Calculations can thus be performed using the known photoionization cross-sections, that are in a field free environment \cite{COM181}
$ \sigma^{\rm pi}_{n' l' m'  \rightarrow E l m  }  = \frac{2 \pi^2 \nu  e^2 }{  \varepsilon_0 c  }  |\langle E l m| \bm r  | n' l' m' \rangle|^2 $. 

Using the Milne's relations we can calculate the cross-sections toward all levels with a given $m'$ quantum number. The results for unpolarized light are given in Fig.~\ref{fig:9a} and \ref{fig:9b}.
\\ \\
The first important result is that the srr process forms low angular momentum states because only small values of $l'$ (or $m'$) contribute to the cross-sections \cite{katkov1978radiative,hahn1997electron}. A second important point is that the sum of the cross-sections over the $n'^2$ ($l',m'$) states within the manifold $n'$ is very close to the Kramer's approximation. In other words, the sum in Fig.~\ref{fig:radiativerecombination} is close to $1$ which is thus a quite accurate formula, especially for low energy and high $n$ states \cite{ROB08,kotelnikov2019electron} (errors have been called Gaunt factors \cite{1988ApJ...327..485R}). Using the fact that $\kappa \ll n'$ in our case, we find a very useful approximation 
$\sigma^{\rm srr, Kramers}_{E= \kappa^2 R_y \rightarrow n'} 
\approx \unit[6.96 \times 10^{-42} ]{m^2} \frac{n'^3}{ \kappa^2}$.
This leads to  
\begin{equation}
    \Gamma^{\rm{srr}} \approx N_1 \times  \unit[36.3]{s^{-1}} \times \left(\frac{n'}{10}\right)^5 
 \frac{I}{ W/m^2}  \frac{\sigma^{\rm srr}}{\sigma^{\rm srr, Kramers}_{E= \kappa^2 R_y \rightarrow n'} }. 
 \label{eq_srr_kramers}  
\end{equation}
The $n'^5$ dependence originates from the $n'^2$ degeneracy and the $n'^3$ dependence of the dipole transition strength in field free. 
This simple expression can be combined with (almost triangular shape) results in Fig.~\ref{fig:radiativerecombination} to provide very simple estimations:
$$\frac{\sigma^{\rm srr}}{\sigma^{\rm srr, Kramers}_{E= \kappa^2 R_y \rightarrow n'} } \approx \frac{1}{m'_{\rm max}}- \frac{ |m'|}{{m'}_{\rm max}^2}.$$
This formula is valid for $|m'| \leq m'_{\rm max} = 3 n^{0.7}/2$ whereas $\sigma^{\rm srr}$ is almost zero elsewhere.

\bibliographystyle{unsrt}
\bibliography{Bibliography_Apr2020}

\end{document}